# Reaching Optimized Parameter Set:

# Protein Secondary Structure Prediction Using Neural Network

DongardiveJyotshna*[a], Siby Abraham[*b]

[a]*Department of Computer Science, University of Mumbai, Mumbai, India* ;

[b]*Department of Mathematics and Statistics, G.N. Khalsa College, University of Mumbai, India.*

Email addresses:

[a.] jyotss.d@gmail.com

[b.] sibyam@gmail.com

**Abstract:** We propose an optimized parameter set for protein secondary structure prediction using three layer feed forward back propagation neural network. The methodology uses four parameters viz. encoding scheme, window size, number of neurons in the hidden layer and type of learning algorithm. The input layer of the network consists of neurons changing from 3 to 19, corresponding to different window sizes. The hidden layer chooses a natural number from 1 to 20 as the number of neurons. The output layer consists of three neurons, each corresponding to known secondary structural classes viz. α – helix, β-strands and coil/turns respectively. It also uses eight different learning algorithms and nine encoding schemes. Exhaustive experiments were performed using non-homologues dataset. The experimental results were compared using performance measures like $Q_3$, sensitivity, specificity, Mathew correlation coefficient and accuracy. The paper also discusses the process of obtaining a stabilized cluster of 2530 records from a collection of 11340 records. The graphs of these stabilized clusters of records with respect to accuracy are concave, convergence is monotonic increasing and rate of convergence is uniform. The paper gives BLOSUM62 as the encoding scheme, 19 as the window size, 19 as the number of neurons in the hidden layer and One- Step Secant as the learning algorithm with the highest accuracy of 78%. These parameter values are proposed as the optimized parameter set for the three layer feed forward back propagation neural network for the protein secondary structure prediction.

## 1. INTRODUCTION

Proteins are made up of simple building blocks called amino acids, which consist of a carbon atom to which a primary amino group, a carboxylic acid group, a side chain (R group) and an H atom are attached as shown in Figure 1. There are numerous amino acids in nature, but only 20 are proteinogenic.

### 1.1. Proteins and their structures

Proteins are organised in four different structural levels [28] [3] [90] [102]. They are primary, secondary, tertiary and quaternary structures. Primary structure (1-D) refers to the amino acid sequence of a protein. It provides foundation of all other types of structures. Secondary structure (2-D) refers to the arrangement of connections within the amino acid groups to form local structures. α –helix (H) [104], β-strands (E) [105] and coil/turns (C) [89] are examples of these. Tertiary structure (3-D) is the three dimensional folding of secondary structures of a polypeptide chain. Quaternary structure (4-D) is formed from interactions of several independent polypeptide chains. The four structures of proteins are shown in Figure 2.

The paper, which deals with the prediction of protein secondary structure, is described in four sections. The remaining part of the section 1 deals with different online databases, different techniques of secondary structure prediction and multilayer feed forward neural network. Section 2 describes materials and methods used in the work. Section 3 discusses the experimental results and their analysis. Section 4 gives the conclusion.

### 1.2. Protein database

Protein data is obtained by experimental approaches like X-ray [39] [128], Nuclear Magnetic Resonance (NMR) [74] [135] and Electron Microscopy (EM). This data are stored in different databases based on their characteristics. Such databases range from simple sequence repositories to curated databases. The simple sequence repositories store data with little or no manual intervention in the creation of the records, while curated databases store annotated sequence data. The various types of online protein databases and their details are shown in Table 1.

Table **1**. Lists some protein databases.

| **Database** | **Structure** | **Description** | **References** |
|---|---|---|---|
| UniProtKB/TrEMBL | Primary | Repository of protein amino acid sequences consist of name/description, taxonomic data and citation information | [5][129][130] |
| PIR | Secondary | A curated database of protein sequence alignments. | [126][11][69][95] |
| eMOTIF | Secondary | Contains Protein sequence motif determination and searches | [70] |
| PROSITE | Primary/Secondary | Describe sequence motif definitions, protein domains, families and functional patterns | [64] |
| PRINTS | Primary/Secondary | A compendium of protein fingerprints. A fingerprint is a group of conserved motifs used to characterise a protein family. | [8][7] |
| BLOCKS | Primary/Secondary | Contains Multiple-alignment of conserved regions of protein families. | [60][57][56][58] |
| INTERPRO | Secondary/Tertiary | Repository of Protein families and domain | [4] |
| PDB | Tertiary | Contains Structure data determined by X-ray crystallography and NMR | [17][15] |
| PRODOM | Tertiary | Repository of Protein domain Families | [29][30] |

### 1.3. Prediction

Despite the growing number of protein sequences, only a few of them have their secondary and tertiary structures unveiled. For instance, the UniProtKB/TrEMBL [130] repository of protein sequences has currently around 50825784 sequence entries (as on 16$^{th}$ September 2015), and the Protein Data Bank (PDB) registers [16] the structure of only 112131 proteins (as on 16$^{th}$ September 2015). From biochemical and biological point of view, this shortage in the number of known protein structures with respect to the number of known sequences is due to the cost and difficulty in unveiling the structures.

Finding the proteins that make up an organism (which is referred as the protein folding problem in bioinformatics) and understanding their functions is the foundation of molecular biology [72]. It is through the tertiary structure of the proteins that we can derive its properties as well as how they function in an organism. Secondary structure prediction, in which secondary structure is predicted from its primary sequences, is an essential intermediate step in the prediction of 3-D structures.

Different techniques have been developed to predict secondary structure of proteins from their primary sequences. Some of the computational methods that are used to achieve secondary structure predictions include Artificial Neural Networks(ANN) [20] [21] [111] [65] [81] [62] [91] [127][140][115][116],Support Vector Machines [31] [79] [68][85],Statistical methods [26] [27] [101] [48] [94][49] [43][80] and Nearest Neighbor Methods [120]. These computational techniques try to overcome the difficulties faced in the biochemical and biological approaches of protein secondary structure prediction. Of these, artificial neural network is the most often used method. [50][82][134] [20][21][111][65][77] [117] [109]. A review of literature on computational techniques for secondary structure prediction using neural network indicates that multilayer feed forward neural networks are the most preferred and effective tool [111][118][19].

### 1.4. Multilayer feed forward neural network in secondary structure prediction

A multilayer feed forward neural network consists of one input layer, one output layer and at least one hidden layer. These layers are interconnected as shown in Figure 3. The number of units, known as neurons, in each layer depends upon the problem under study. Each unit in the input layer supplies a signal to every unit in the first hidden layer. The output, which is transformed by a transformation function, is passed to units in the next hidden layer or the units in the output layer depending upon the number of hidden layers [118] [10]. Thus, the connected units form a network. Each connection between units has a weight attached to it. The amount of change in the network is determined according to an error correction-learning algorithm. The network is trained to create an input-output model with correct mapping such that for unseen inputs, their outputs can be predicted successfully [92]. There are many approaches in a multi layer feed forward neural networks. However, multilayer feed forward back propagation networks are the most efficient ones [18].

Back propagation learning technique is a supervised learning technique in which all units in different layers undergo two passes viz. forward pass and backward pass. During forward pass, all synaptic weights are fixed and a signal given to each unit in the input layer is propagated layer by layer until it reaches the unit in the output layer. This actual output is compared with the expected output and the difference, known as error, is propagated back. During backward pass, synaptic weights are adjusted according to an error correction rule [55], the most often used one is the generalized delta rule. This process is continued iteratively through a series of forward and backward passes until the network gives the desired response as the output. The process is continued for a number of input-output pairs to train the network. A typical multi layer feed forward back propagation network with single hidden layer for secondary structure prediction assumes a number of parameters. These parameters are data encoding scheme, window size, number of hidden neurons and type of learning algorithms.

The encoding scheme arranges the input data in a format, which can be passed to the neural network. However, all input data cannot directly be supplied to the network. It has to be structured into parts, which can be achieved using sliding window protocol. By assuming a single hidden layer in the network, the number of neurons in that hidden layer affects the performance of the neural network substantially. The training of the network, which is the most important aspect in the neural network, can be realized using different tried and tested learning algorithms.

A survey on literature of secondary structure prediction using multi-layer feed forward back propagation neural networks shows that the highest accuracy is obtained around 64.3% (with respect to performance measure $Q_3$) with a small variation caused by datasets used [111][65][37][22][63]. There have been attempts to increase the value of performance by using pre-processing strategies [124][78],incorporating domain specific heuristic information [87] [24] [97] [25] [115] [116] [123] [131] [32] [108][88][40][22][96][137][41] and using hybridization of neural network with other computational techniques [76][71][12][133][6][61][138][84][139][86].

The proposed work does not use any of these fine-tuning techniques, as the objective is to find the best parameter set on a conventional setup. Related works attempt to predict secondary structure usually by changing one or two parameters [111] [65] [98][47][115] [116][114] [25] [136] [9] [100] [54] [35] [2] [1] [75] [38]. The authors, to the best of their knowledge, are yet to find a work, which considers changes in more than three parameters. The method proposed in this work considers changes in all the four parameters. This uses nine data encoding schemes (ES), window size (WS) ranges from three to nineteen (nine in number), twenty different hidden neurons (HN) and eight learning algorithms (LA). It offers an optimized parameter set by exhaustive search in a search space of 9*9*20*8=12960 search points. This has been validated by experimental results using five different performance measures. Other related works vary only one or two performance measures and with problems of lesser search complexity. The optimized parameter set proposed also incorporates the distinct behaviour determined by these five different performance measures.

## 2. MATERIALS AND METHODS

The methodology proposed to predict the secondary structure of protein is shown in Figure 4. The different stages starting from collection of data to calculation of performance measures is discussed in following sections.

### 2.1. Data collection

The data used in the study is RS126, which is one of the oldest dataset used for protein secondary structure prediction. The scheme, which is created by Rost and Sander [114] consists of 126 sequences of average sequence length 186 and 23,347 residues. RS126 dataset is collected from supplementary data files in previous research or study. Besides, it can also be obtained from online databases such as PDB.

The data collected is structured in rows by protein name, primary and secondary structures. The primary structure is a sequence of amino acids, which are represented by one letter code. The secondary structure of proteins is represented in three structural classes namely α−helices, β−strands and coil/turns and the rest are represented with a dash (−) as used by Cuff and Barton [33] and Hua and Sun [68]. A sample of the data used is shown in Table 2.

Table **2**. Sample data used.

| | |
|---|---|
| Protein Name | 1CBH:A |
| Primary Sequence | TQSHYGQCGGIGYSGPTVCASGTTCQV LNPYYSQCL |
| Secondary Structure | ---CC-EEE-CC--C-----CC--EEEECCEEEE- |

### 2.2. Data encoding

Data encoding is proposed to convert amino acids, which are represented by single letter code to its numerical equivalent. This is to facilitate the data to be used by the neural network framework. The different encoding schemes used in the work are Orthogonal, Hydrophobicity, BLOSUM62, PAM250 and Hybrid encoding schemes viz. Orthogonal+Hydrophobicity,BLOSUM62 + Hydrophobicity, Orthogonal + BLOSUM62, PAM250 + Hydrophobicity and Orthogonal + PAM250. Each of the schemes offers a matrix representation for the given primary sequence with the number of rows corresponding to the length of the sequence (number of amino acids in the sequence) and the number of columns corresponding to 20, the number of different amino acids. The schemes vary according to the way the entry in the matrix is calculated. The following subsections give a brief discussion on each of these encoding schemes as implemented in the work.

#### 2.2.1 Orthogonal encoding

The orthogonal encoding scheme, suggested by Holley and Karplus [65], uses binary digits 0 and 1 to represent an amino acid. For a given row in the matrix, the presence of an amino acid is represented as 1 and all other entries are marked as 0. For example, the amino acid H which appears in the seventh position in the sequence as given in EQ.1

ACDEFGHIKLMNPQRSTVWY (EQ.1)

is encoded as

$$(00000010000000000000)^{T} \quad \text{(EQ. 2)}$$

and is represented as the seventh row in the matrix as shown in Table 3.

Table **3**. Orthogonal encoding

| | A | C | D | E | F | G | H | I | K | L | M | N | P | Q | R | S | T | V | W | Y |
|---|---|---|---|---|---|---|---|---|---|---|---|---|---|---|---|---|---|---|---|---|
| A | 1 | 0 | 0 | 0 | 0 | 0 | 0 | 0 | 0 | 0 | 0 | 0 | 0 | 0 | 0 | 0 | 0 | 0 | 0 | 0 |
| C | 0 | 1 | 0 | 0 | 0 | 0 | 0 | 0 | 0 | 0 | 0 | 0 | 0 | 0 | 0 | 0 | 0 | 0 | 0 | 0 |
| D | 0 | 0 | 1 | 0 | 0 | 0 | 0 | 0 | 0 | 0 | 0 | 0 | 0 | 0 | 0 | 0 | 0 | 0 | 0 | 0 |
| E | 0 | 0 | 0 | 1 | 0 | 0 | 0 | 0 | 0 | 0 | 0 | 0 | 0 | 0 | 0 | 0 | 0 | 0 | 0 | 0 |
| F | 0 | 0 | 0 | 0 | 1 | 0 | 0 | 0 | 0 | 0 | 0 | 0 | 0 | 0 | 0 | 0 | 0 | 0 | 0 | 0 |
| G | 0 | 0 | 0 | 0 | 0 | 1 | 0 | 0 | 0 | 0 | 0 | 0 | 0 | 0 | 0 | 0 | 0 | 0 | 0 | 0 |
| H | 0 | 0 | 0 | 0 | 0 | 0 | 1 | 0 | 0 | 0 | 0 | 0 | 0 | 0 | 0 | 0 | 0 | 0 | 0 | 0 |
| I | 0 | 0 | 0 | 0 | 0 | 0 | 0 | 1 | 0 | 0 | 0 | 0 | 0 | 0 | 0 | 0 | 0 | 0 | 0 | 0 |
| K | 0 | 0 | 0 | 0 | 0 | 0 | 0 | 0 | 1 | 0 | 0 | 0 | 0 | 0 | 0 | 0 | 0 | 0 | 0 | 0 |
| L | 0 | 0 | 0 | 0 | 0 | 0 | 0 | 0 | 0 | 1 | 0 | 0 | 0 | 0 | 0 | 0 | 0 | 0 | 0 | 0 |
| M | 0 | 0 | 0 | 0 | 0 | 0 | 0 | 0 | 0 | 0 | 1 | 0 | 0 | 0 | 0 | 0 | 0 | 0 | 0 | 0 |
| N | 0 | 0 | 0 | 0 | 0 | 0 | 0 | 0 | 0 | 0 | 0 | 1 | 0 | 0 | 0 | 0 | 0 | 0 | 0 | 0 |
| P | 0 | 0 | 0 | 0 | 0 | 0 | 0 | 0 | 0 | 0 | 0 | 0 | 1 | 0 | 0 | 0 | 0 | 0 | 0 | 0 |
| Q | 0 | 0 | 0 | 0 | 0 | 0 | 0 | 0 | 0 | 0 | 0 | 0 | 0 | 1 | 0 | 0 | 0 | 0 | 0 | 0 |
| R | 0 | 0 | 0 | 0 | 0 | 0 | 0 | 0 | 0 | 0 | 0 | 0 | 0 | 0 | 1 | 0 | 0 | 0 | 0 | 0 |
| S | 0 | 0 | 0 | 0 | 0 | 0 | 0 | 0 | 0 | 0 | 0 | 0 | 0 | 0 | 0 | 1 | 0 | 0 | 0 | 0 |
| T | 0 | 0 | 0 | 0 | 0 | 0 | 0 | 0 | 0 | 0 | 0 | 0 | 0 | 0 | 0 | 0 | 1 | 0 | 0 | 0 |
| V | 0 | 0 | 0 | 0 | 0 | 0 | 0 | 0 | 0 | 0 | 0 | 0 | 0 | 0 | 0 | 0 | 0 | 1 | 0 | 0 |
| W | 0 | 0 | 0 | 0 | 0 | 0 | 0 | 0 | 0 | 0 | 0 | 0 | 0 | 0 | 0 | 0 | 0 | 0 | 1 | 0 |
| Y | 0 | 0 | 0 | 0 | 0 | 0 | 0 | 0 | 0 | 0 | 0 | 0 | 0 | 0 | 0 | 0 | 0 | 0 | 0 | 1 |

### 2.2.2 Hydrophobicity encoding

Hydrophobicity encoding scheme, suggested by Radzicka and Wolfenden [112], uses the hydrophobicity index (hi), as given in Table 4, for each of the amino acids. According to this, a hydrophobicity matrix (hm) is created wherein the entries are calculated by the formula given in EQ. 3.

$$hm\,[i]\,[j] = \frac{abs\,(hi[i]-hi[j])}{20.0} \quad \text{(EQ.3)}$$

Table **4**. Hydrophobicity index for each of the amino acids

| Amino acids (.) | Hydrophobicity index {.} |
|---|---|
| A | 1.81 |
| R | -14.92 |
| N | -6.64 |
| D | -8.72 |
| C | 1.28 |
| Q | -5.54 |
| E | -6.81 |
| G | 0.94 |
| H | -4.66 |
| I | 4.92 |
| L | 4.92 |
| K | -5.55 |
| M | 2.35 |
| F | 2.98 |
| P | 4.04 |
| S | -3.40 |
| T | -2.57 |
| W | 2.33 |
| Y | -0.14 |
| V | 4.04 |

For example the amino acids A (1.81) and D (-8.72) in EQ.1 becomes 0.525. Based on this method, a 20 by 20 hydrophobicity matrix for the sequence given in (EQ. 1) is formulated as shown in Table 5.

Table **5**. Hydrophobicity encoding

| | A | R | N | D | C | Q | E | G | H | I | L | K | M | F | P | S | T | W | Y | V |
|---|---|---|---|---|---|---|---|---|---|---|---|---|---|---|---|---|---|---|---|---|
| A | 0 | 0.83 | 0.42 | 0.52 | 0.02 | 0.36 | 0.43 | 0.04 | 0.32 | 0.15 | 0.15 | 0.36 | 0.02 | 0.05 | 0.11 | 0.26 | 0.21 | 0.02 | 0.09 | 0.11 |
| R | 0.83 | 0 | 0.41 | 0.31 | 0.81 | 0.46 | 0.40 | 0.79 | 0.51 | 0.99 | 0.99 | 0.46 | 0.86 | 0.89 | 0.94 | 0.57 | 0.61 | 0.86 | 0.73 | 0.94 |
| N | 0.42 | 0.41 | 0 | 0.10 | 0.39 | 0.05 | 0.00 | 0.37 | 0.09 | 0.57 | 0.57 | 0.05 | 0.44 | 0.48 | 0.53 | 0.16 | 0.20 | 0.44 | 0.32 | 0.53 |
| D | 0.52 | 0.31 | 0.10 | 0 | 0.5 | 0.15 | 0.09 | 0.48 | 0.20 | 0.68 | 0.68 | 0.15 | 0.55 | 0.58 | 0.63 | 0.26 | 0.30 | 0.55 | 0.42 | 0.63 |
| C | 0.02 | 0.81 | 0.39 | 0.5 | 0 | 0.34 | 0.40 | 0.01 | 0.29 | 0.18 | 0.18 | 0.34 | 0.05 | 0.08 | 0.13 | 0.23 | 0.19 | 0.05 | 0.07 | 0.13 |
| Q | 0.36 | 0.46 | 0.05 | 0.15 | 0.34 | 0 | 0.06 | 0.32 | 0.04 | 0.52 | 0.52 | 0.00 | 0.39 | 0.42 | 0.47 | 0.10 | 0.14 | 0.39 | 0.27 | 0.47 |
| E | 0.43 | 0.40 | 0.00 | 0.09 | 0.40 | 0.06 | 0 | 0.38 | 0.10 | 0.58 | 0.58 | 0.06 | 0.45 | 0.48 | 0.54 | 0.17 | 0.21 | 0.45 | 0.33 | 0.54 |
| G | 0.04 | 0.79 | 0.37 | 0.48 | 0.01 | 0.32 | 0.38 | 0 | 0.28 | 0.19 | 0.19 | 0.32 | 0.07 | 0.10 | 0.15 | 0.21 | 0.17 | 0.06 | 0.05 | 0.15 |
| H | 0.32 | 0.51 | 0.09 | 0.20 | 0.29 | 0.04 | 0.10 | 0.28 | 0 | 0.47 | 0.47 | 0.04 | 0.35 | 0.38 | 0.43 | 0.06 | 0.10 | 0.34 | 0.22 | 0.43 |
| I | 0.15 | 0.99 | 0.57 | 0.68 | 0.18 | 0.52 | 0.58 | 0.19 | 0.47 | 0 | 0 | 0.52 | 0.12 | 0.09 | 0.04 | 0.41 | 0.37 | 0.12 | 0.25 | 0.04 |
| L | 0.15 | 0.99 | 0.57 | 0.68 | 0.18 | 0.52 | 0.58 | 0.19 | 0.47 | 0 | 0 | 0.52 | 0.12 | 0.09 | 0.04 | 0.41 | 0.37 | 0.12 | 0.25 | 0.04 |
| K | 0.36 | 0.46 | 0.05 | 0.15 | 0.34 | 0.00 | 0.06 | 0.32 | 0.04 | 0.52 | 0.52 | 0 | 0.39 | 0.42 | 0.47 | 0.10 | 0.14 | 0.39 | 0.27 | 0.47 |
| M | 0.02 | 0.86 | 0.44 | 0.55 | 0.05 | 0.39 | 0.45 | 0.07 | 0.35 | 0.12 | 0.12 | 0.39 | 0 | 0.03 | 0.08 | 0.28 | 0.24 | 0.00 | 0.12 | 0.08 |
| F | 0.05 | 0.89 | 0.48 | 0.58 | 0.08 | 0.42 | 0.48 | 0.10 | 0.38 | 0.09 | 0.09 | 0.42 | 0.03 | 0 | 0.05 | 0.31 | 0.27 | 0.03 | 0.15 | 0.05 |
| P | 0.11 | 0.94 | 0.53 | 0.63 | 0.13 | 0.47 | 0.54 | 0.15 | 0.43 | 0.04 | 0.04 | 0.47 | 0.08 | 0.05 | 0 | 0.37 | 0.33 | 0.08 | 0.20 | 0 |
| S | 0.26 | 0.57 | 0.16 | 0.26 | 0.23 | 0.10 | 0.17 | 0.21 | 0.06 | 0.41 | 0.41 | 0.10 | 0.28 | 0.31 | 0.37 | 0 | 0.04 | 0.28 | 0.16 | 0.37 |
| T | 0.21 | 0.61 | 0.20 | 0.30 | 0.19 | 0.14 | 0.21 | 0.17 | 0.10 | 0.37 | 0.37 | 0.14 | 0.24 | 0.27 | 0.33 | 0.04 | 0 | 0.24 | 0.12 | 0.33 |
| W | 0.02 | 0.86 | 0.44 | 0.55 | 0.05 | 0.39 | 0.45 | 0.06 | 0.34 | 0.12 | 0.12 | 0.39 | 0.00 | 0.03 | 0.08 | 0.28 | 0.24 | 0 | 0.12 | 0.08 |
| Y | 0.09 | 0.73 | 0.32 | 0.42 | 0.07 | 0.27 | 0.33 | 0.05 | 0.22 | 0.25 | 0.25 | 0.27 | 0.12 | 0.15 | 0.20 | 0.16 | 0.12 | 0.12 | 0 | 0.20 |

| | | | | | | | | | | | | | | | | | | | |
|---|---|---|---|---|---|---|---|---|---|---|---|---|---|---|---|---|---|---|---|
| V | 0.11 | 0.94 | 0.53 | 0.63 | 0.13 | 0.47 | 0.54 | 0.15 | 0.43 | 0.04 | 0.04 | 0.47 | 0.08 | 0.05 | 0 | 0.37 | 0.33 | 0.08 | 0.20 | 0 |

### 2.2.3 BLOSUM62 encoding

BLOSUM62 substitution matrix [59][67], as shown in Table 6, provides a ‘log- odds’ score for the possibility of a given pair of amino acids interchanging with each other. BLOSUM62 encoding scheme uses this substitution matrix for the representation of amino acid in the sequence.

Table **6.** BLOSUM62 substitution matrix

| | A | R | N | D | C | Q | E | G | H | I | L | K | M | F | P | S | T | W | Y | V |
|---|---|---|---|---|---|---|---|---|---|---|---|---|---|---|---|---|---|---|---|---|
| A | 4 | -1 | -2 | -2 | 0 | -1 | -1 | 0 | -2 | -1 | -1 | -1 | -1 | -2 | -1 | 1 | 0 | -3 | -2 | 0 |
| R | 0 | -3 | -3 | -3 | 9 | -3 | -4 | -3 | -3 | -1 | -1 | -3 | -1 | -2 | -3 | -1 | -1 | -2 | -2 | -1 |
| N | -2 | -2 | 1 | 6 | -3 | 0 | 2 | -1 | -1 | -3 | -4 | -1 | -3 | -3 | -1 | 0 | -1 | -4 | -3 | -3 |
| D | -1 | 0 | 0 | 2 | -4 | 2 | 5 | -2 | 0 | -3 | -3 | 1 | -2 | -3 | -1 | 0 | -1 | -3 | -2 | -2 |
| C | -2 | -3 | -3 | -3 | -2 | -3 | -3 | -3 | -1 | 0 | 0 | -3 | 0 | 6 | -4 | -2 | -2 | 1 | 3 | -1 |
| Q | 0 | -2 | 0 | -1 | -3 | -2 | -2 | 6 | -2 | -4 | -4 | -2 | -3 | -3 | -2 | 0 | -2 | -2 | -3 | -3 |
| E | -2 | 0 | 1 | -1 | -3 | 0 | 0 | -2 | 8 | -3 | -3 | -1 | -2 | -1 | -2 | -1 | -2 | -2 | 2 | -3 |
| G | -1 | -3 | -3 | -3 | -1 | -3 | -3 | -4 | -3 | 4 | 2 | -3 | 1 | 0 | -3 | -2 | -1 | -3 | -1 | 3 |
| H | -1 | 2 | 0 | -1 | -3 | 1 | 1 | -2 | -2 | -3 | -2 | 5 | -1 | -3 | -1 | 0 | -1 | -3 | -2 | -2 |
| I | -1 | -2 | -3 | -4 | -1 | -2 | -3 | -4 | -3 | 2 | 4 | -2 | 2 | 0 | -3 | -2 | -1 | -2 | -1 | 1 |
| L | -1 | -1 | -2 | -3 | -1 | 0 | -2 | -3 | -2 | 1 | 2 | -1 | 5 | 0 | -2 | -1 | -1 | -1 | -1 | 1 |
| K | -2 | 0 | 6 | 1 | -3 | 0 | -3 | 1 | -3 | 0 | -3 | 0 | -2 | 0 | -2 | 1 | 0 | -4 | -2 | -3 |
| M | -1 | -2 | -2 | -1 | -3 | -1 | -1 | -2 | -2 | -3 | -3 | -1 | -2 | -4 | 7 | -1 | -1 | -4 | -3 | -2 |
| F | -1 | 1 | 0 | 0 | -3 | 5 | 2 | -2 | 0 | -3 | -2 | 1 | 0 | -3 | -1 | -1 | -1 | -2 | -2 | -1 |
| P | -1 | 5 | 0 | -2 | -3 | 1 | 0 | -2 | 0 | -3 | -2 | 2 | -1 | -3 | -2 | -1 | -1 | -3 | -2 | -3 |
| S | 1 | -1 | 1 | 0 | -1 | 0 | 0 | 0 | -1 | -2 | -2 | 0 | -1 | -2 | -1 | 4 | 1 | -3 | -2 | -2 |
| T | 0 | -1 | 0 | -1 | -1 | -1 | -1 | -2 | -2 | -1 | -1 | -1 | -1 | -2 | -1 | 1 | 5 | -2 | -2 | 0 |
| W | 0 | -3 | -3 | -3 | -1 | -2 | -2 | -3 | -3 | 3 | 1 | -2 | 1 | -1 | -2 | -2 | 0 | -3 | -1 | 4 |
| Y | -3 | -3 | -4 | -4 | -2 | -2 | -3 | -2 | -2 | -3 | -2 | -3 | -1 | 1 | -4 | -3 | -2 | 11 | 2 | -3 |
| V | -2 | -2 | -2 | -3 | -2 | -1 | -2 | -3 | 2 | -1 | -1 | -2 | -1 | 3 | -3 | -2 | -2 | 2 | 7 | -1 |

Accordingly, for each amino acid, the corresponding row from the BLOSUM62 substitution matrix is identified and placed as the representation for that amino acid in the encoded matrix. For example, the amino acid G in the sequence (EQ. 1) is represented as

$$(0\ -2\ 0\ -1\ -3\ -2\ -2\ 6\ -2\ -4\ -4\ -2\ -3\ -3\ -2\ 0\ -2\ -2\ -3)^T \quad \text{(EQ.4)}$$

as shown in Table 7.

Table **7**. BLOSUM62 substitution matrix encoded for the sequence

| | A | C | D | E | F | G | H | I | K | L | M | N | P | Q | R | S | T | V | W | Y |
|---|---|---|---|---|---|---|---|---|---|---|---|---|---|---|---|---|---|---|---|---|
| A | 4 | 0 | -2 | -1 | -2 | 0 | -2 | -1 | -1 | -1 | -1 | -2 | -1 | -1 | -1 | 1 | 0 | 0 | -3 | -2 |
| C | 0 | 9 | -3 | -4 | -2 | -3 | -3 | -1 | -3 | -1 | -1 | -3 | -3 | -3 | -3 | -1 | -1 | -1 | -2 | -2 |
| D | -2 | -3 | 6 | 2 | -3 | -1 | -1 | -3 | -1 | -4 | -3 | 1 | -1 | 0 | -2 | 0 | -1 | -3 | -4 | -3 |
| E | -1 | -4 | 2 | 5 | -3 | -2 | 0 | -3 | 1 | -3 | -2 | 0 | -1 | 2 | 0 | 0 | -1 | -2 | -3 | -2 |
| F | -2 | -2 | -3 | -3 | 6 | -3 | -1 | 0 | -3 | 0 | 0 | -3 | -4 | -3 | -3 | -2 | -2 | -1 | 1 | 3 |
| G | 0 | -3 | -1 | -2 | -3 | 6 | -2 | -4 | -2 | -4 | -3 | 0 | -2 | -2 | -2 | 0 | -2 | -3 | -2 | -3 |
| H | -2 | -3 | -1 | 0 | -1 | -2 | 8 | -3 | -1 | -3 | -2 | 1 | -2 | 0 | 0 | -1 | -2 | -3 | -2 | 2 |
| I | -1 | -1 | -3 | -3 | 0 | -4 | -3 | 4 | -3 | 2 | 1 | -3 | -3 | -3 | -3 | -2 | -1 | 3 | -3 | -1 |
| K | -1 | -3 | -1 | 1 | -3 | -2 | -2 | -3 | 5 | -2 | -1 | 0 | -1 | 1 | 2 | 0 | -1 | -2 | -3 | -2 |
| L | -1 | -1 | -4 | -3 | 0 | -4 | -3 | 2 | -2 | 4 | 2 | -3 | -3 | -2 | -2 | -2 | -1 | 1 | -2 | -1 |
| M | -1 | -1 | -3 | -2 | 0 | -3 | -2 | 1 | -1 | 2 | 5 | -2 | -2 | 0 | -1 | -1 | -1 | 1 | -1 | -1 |
| N | -2 | -3 | 1 | -3 | 0 | 1 | -3 | 0 | 0 | -3 | -2 | 6 | -2 | 0 | 0 | 1 | 0 | -3 | -4 | -2 |
| P | -1 | -3 | -1 | -1 | -4 | -2 | -2 | -3 | -1 | -3 | -2 | -2 | 7 | -1 | -2 | -1 | -1 | -2 | -4 | -3 |
| Q | -1 | -3 | 0 | 2 | -3 | -2 | 0 | -3 | 1 | -2 | 0 | 0 | -1 | 5 | 1 | -1 | -1 | -1 | -2 | -2 |
| R | -1 | -3 | -2 | 0 | -3 | -2 | 0 | -3 | 2 | -2 | -1 | 0 | -2 | 1 | 5 | -1 | -1 | -3 | -3 | -2 |
| S | 1 | -1 | 0 | 0 | -2 | 0 | -1 | -2 | 0 | -2 | -1 | 1 | -1 | 0 | -1 | 4 | 1 | -2 | -3 | -2 |
| T | 0 | -1 | -1 | -1 | -2 | -2 | -2 | -1 | -1 | -1 | -1 | 0 | -1 | -1 | -1 | 1 | 5 | 0 | -2 | -2 |
| V | 0 | -1 | -3 | -2 | -1 | -3 | -3 | 3 | -2 | 1 | 1 | -3 | -2 | -2 | -3 | -2 | 0 | 4 | -3 | -1 |
| W | -3 | -2 | -4 | -3 | 1 | -2 | -2 | -3 | -3 | -2 | -1 | -4 | -4 | -2 | -3 | -3 | -2 | -3 | 11 | 2 |
| Y | -2 | -2 | -3 | -2 | 3 | -3 | 2 | -1 | -2 | -1 | -1 | -2 | -3 | -1 | -2 | -2 | -2 | -1 | 2 | 7 |

### 2.2.4 PAM250 encoding

PAM250 Mutation matrix [34], as shown in Table 8, provides the number of mutations taking place for each amino acid over an evolutionary distance.

Table **8**. PAM250 mutation matrix

| | A | R | N | D | C | Q | E | G | H | I | L | K | M | F | P | S | T | W | Y | V |
|---|---|---|---|---|---|---|---|---|---|---|---|---|---|---|---|---|---|---|---|---|
| A | 2 | -2 | 0 | 0 | -2 | 0 | 0 | 1 | -1 | -1 | -2 | -1 | -1 | -3 | 1 | 1 | 1 | -6 | -3 | 0 |
| R | -2 | 6 | 0 | -1 | -4 | 1 | -1 | -3 | 2 | -2 | -3 | 3 | 0 | -4 | 0 | 0 | -1 | 2 | -4 | -2 |
| N | 0 | 0 | 2 | 2 | -4 | 1 | 1 | 0 | 2 | -2 | -3 | 1 | -2 | -3 | 0 | 1 | 0 | -4 | -2 | -2 |
| D | 0 | -1 | 2 | 4 | -5 | 2 | 3 | 1 | 1 | -2 | -4 | 0 | -3 | -6 | -1 | 0 | 0 | -7 | -4 | -2 |
| C | -2 | -4 | -4 | -5 | 12 | -5 | -5 | -3 | -3 | -2 | -6 | -5 | -5 | -4 | -3 | 0 | -2 | -8 | 0 | -2 |
| Q | 0 | 1 | 1 | 2 | -5 | 4 | 2 | -1 | 3 | -2 | -2 | 1 | -1 | -5 | 0 | -1 | -1 | -5 | -4 | -2 |
| E | 0 | -1 | 1 | 3 | -5 | 2 | 4 | 0 | 1 | -2 | -3 | 0 | -2 | -5 | -1 | 0 | 0 | -7 | -4 | -2 |
| G | 1 | -3 | 0 | 1 | -3 | -1 | 0 | 5 | -2 | -3 | -4 | -2 | -3 | -5 | 0 | 1 | 0 | -7 | -5 | -1 |
| H | -1 | 2 | 2 | 1 | -3 | 3 | 1 | -2 | 6 | -2 | -2 | 0 | -2 | -2 | 0 | -1 | -1 | -3 | 0 | -2 |
| I | -1 | -2 | -2 | -2 | -2 | -2 | -2 | -3 | -2 | 5 | 2 | -2 | 2 | 1 | -2 | -1 | 0 | -5 | -1 | 4 |
| L | -2 | -3 | -3 | -4 | -6 | -2 | -3 | -4 | -2 | 2 | 6 | -3 | 4 | 2 | -3 | -3 | -2 | -2 | -1 | 2 |
| K | -1 | 3 | 1 | 0 | -5 | -1 | 0 | -2 | 0 | -2 | -3 | 5 | 0 | -5 | -1 | 0 | 0 | -3 | -4 | -2 |
| M | -1 | 0 | -2 | -3 | -5 | -1 | -2 | -3 | -2 | 2 | 4 | 0 | 6 | 0 | -2 | -2 | -1 | -4 | -2 | 2 |
| F | -3 | -4 | -3 | -6 | -4 | -5 | -5 | -5 | -2 | 1 | 2 | -5 | 0 | 9 | -5 | -3 | -3 | 0 | 7 | -1 |
| P | 1 | 0 | 0 | -1 | -3 | 0 | -1 | 0 | 0 | -2 | -3 | -1 | -2 | -5 | 6 | 1 | 0 | -6 | -5 | -1 |
| S | 1 | 0 | 1 | 0 | 0 | -1 | 0 | 1 | -1 | -1 | -3 | 0 | -2 | -3 | 1 | 2 | 1 | -2 | -3 | -1 |
| T | 1 | -1 | 0 | 0 | -2 | -1 | 0 | 0 | -1 | 0 | -2 | 0 | -1 | -3 | 0 | 1 | 3 | -5 | -3 | 0 |
| W | -6 | 2 | -4 | -7 | -8 | -5 | -7 | -7 | -3 | -5 | -2 | -3 | -4 | 0 | -6 | -2 | -5 | 17 | 0 | -6 |
| Y | -3 | -4 | -2 | -4 | 0 | -4 | -4 | -5 | 0 | -1 | -1 | -4 | -2 | 7 | -5 | -3 | -3 | 0 | 10 | -2 |
| V | 0 | -2 | -2 | -2 | -2 | -2 | -2 | -1 | -2 | 4 | 2 | -2 | 2 | -1 | -1 | -`1 | 0 | -6 | -2 | 4 |

PAM250 encoding scheme uses this mutation matrix for representation of amino acids and follows the same strategy used in the BLOSUM62 encoding scheme. For example, the amino acid G in the sequence described in (EQ1) is replaced as

$$(1\ -3\ 0\ 1\ -3\ -1\ 0\ 5\ -2\ -3\ -4\ -2\ -3\ -5\ 0\ 1\ 0\ -7\ -5\ -1)^T \qquad \text{(EQ.5)}$$

as shown in Table 9.

Table **9**. PAM250 substitution matrix encoded for the sequence

| | A | C | D | E | F | G | H | I | K | L | M | N | P | Q | R | S | T | V | W | Y |
|---|---|---|---|---|---|---|---|---|---|---|---|---|---|---|---|---|---|---|---|---|
| A | 2 | -2 | 0 | 0 | -3 | 1 | -1 | -1 | -1 | -2 | -1 | 0 | 1 | 0 | -2 | 1 | 1 | 0 | -6 | -3 |
| R | -2 | -4 | -1 | -1 | -4 | -3 | 2 | -2 | 3 | -3 | 0 | 0 | 0 | 1 | 6 | 0 | -1 | -2 | 2 | -4 |
| N | 0 | -4 | 2 | 1 | -3 | 0 | 2 | -2 | 1 | -3 | -2 | 2 | 0 | 1 | 0 | 1 | 0 | -2 | -4 | -2 |
| D | 0 | -5 | 4 | 3 | -6 | 1 | 1 | -2 | 0 | -4 | -3 | 2 | -1 | 2 | -1 | 0 | 0 | -2 | -7 | -4 |
| C | -2 | 12 | -5 | -5 | -4 | -3 | -3 | -2 | -5 | -6 | -5 | -4 | -3 | -5 | -4 | 0 | -2 | -2 | -8 | 0 |
| Q | 0 | -5 | 2 | 2 | -5 | -1 | 3 | -2 | 1 | -2 | -1 | 1 | 0 | 4 | 1 | -1 | -1 | -2 | -5 | -4 |
| E | 0 | -5 | 3 | 4 | -5 | 0 | 1 | -2 | 0 | -3 | -2 | 1 | -1 | 2 | -1 | 0 | 0 | -2 | -7 | -4 |
| G | 1 | -3 | 1 | 0 | -5 | 5 | -2 | -3 | -2 | -4 | -3 | 0 | 0 | -1 | -3 | 1 | 0 | -1 | -7 | -5 |
| H | -1 | -3 | 1 | 1 | -2 | -2 | 6 | -2 | 0 | -2 | -2 | 2 | 0 | 3 | 2 | -1 | -1 | -2 | -3 | 0 |
| I | -1 | -2 | -2 | -2 | 1 | -3 | -2 | 5 | -2 | 2 | 2 | -2 | -2 | -2 | -2 | -1 | 0 | 4 | -5 | -1 |
| L | -2 | -6 | -4 | -3 | 2 | -4 | -2 | 2 | -3 | 6 | 4 | -3 | -3 | -2 | -3 | -3 | -2 | 2 | -2 | -1 |
| K | -1 | -5 | 0 | 0 | -5 | -2 | 0 | -2 | 5 | -3 | 0 | 1 | -1 | -1 | 3 | 0 | 0 | -2 | -3 | -4 |
| M | -1 | -5 | -3 | -2 | 0 | -3 | -2 | 2 | 0 | 4 | 6 | -2 | -2 | -1 | 0 | -2 | -1 | 2 | -4 | -2 |
| F | -3 | -4 | -6 | -5 | 9 | -5 | -2 | 1 | -5 | 2 | 0 | -3 | -5 | -5 | -4 | -3 | -3 | -1 | 0 | 7 |
| P | 1 | -3 | -1 | -1 | -5 | 0 | 0 | -2 | -1 | -3 | -2 | 0 | 6 | 0 | 0 | 1 | 0 | -1 | -6 | -5 |
| S | 1 | 0 | 0 | 0 | -3 | 1 | -1 | -1 | 0 | -3 | -2 | 1 | 1 | -1 | 0 | 2 | 1 | -1 | -2 | -3 |
| T | 1 | -2 | 0 | 0 | -3 | 0 | -1 | 0 | 0 | -2 | -1 | 0 | 0 | -1 | -1 | 1 | 3 | 0 | -5 | -3 |
| W | -6 | -8 | -7 | -7 | 0 | -7 | -3 | -5 | -3 | -2 | -4 | -4 | -6 | -5 | 2 | -2 | -5 | -6 | 17 | 0 |
| Y | -3 | 0 | -4 | -4 | 7 | -5 | 0 | -1 | -4 | -1 | -2 | -2 | -5 | -4 | -4 | -3 | -3 | -2 | 0 | 10 |
| V | 0 | -2 | -2 | -2 | -1 | -1 | -2 | 4 | -2 | 2 | 2 | -2 | -1 | -2 | -2 | -`1 | 0 | 4 | -6 | -2 |

### 2.2.5 Hybrid encoding schemes

Each of the four encoding schemes discussed above captures certain characteristics of amino acids present in the given sequence and is represented as a matrix. In order to have more features of the amino acids to be incorporated, we propose the hybrid encoding of the above encoding schemes by adding the corresponding matrices as suggested by Hu et al [67]. Though the four encoding schemes can offer $4C_2$=6 hybrid encoding schemes by choosing combinations of two schemes, the method avoids the hybrid of PAM250 and BLOSUM62 as they are inappropriate because of their inherent biological nature. Thus, in addition to the above four encoding schemes, the method proposes the following five hybrid encoding schemes also.

- Orthogonal + Hydrophobicity
- BLOSUM62 + Hydrophobicity
- Orthogonal + BLOSUM62
- PAM250 + Hydrophobicity
- Orthogonal + PAM250

The matrices corresponding to these hybrid encodings schemes for the sequence in EQ. 1 are shown in Table 10, Table 11, Table 12, Table 13 and Table 14 respectively.

Table **10**. Orthogonal+Hydrophobicity

| | A | C | D | E | F | G | H | I | K | L | M | N | P | Q | R | S | T | V | W | Y |
|---|---|---|---|---|---|---|---|---|---|---|---|---|---|---|---|---|---|---|---|---|
| A | 1 | 0.02 | 0.52 | 0.43 | 0.05 | 0.04 | 0.32 | 0.15 | 0.36 | 0.15 | 0.02 | 0.42 | 0.11 | 0.36 | 0.83 | 0.26 | 0.21 | 0.11 | 0.02 | 0.09 |
| C | 0.02 | 1 | 0.5 | 0.40 | 0.08 | 0.01 | 0.29 | 0.18 | 0.34 | 0.18 | 0.05 | 0.39 | 0.13 | 0.34 | 0.81 | 0.23 | 0.19 | 0.13 | 0.05 | 0.07 |
| D | 0.52 | 0.5 | 1 | 0.09 | 0.58 | 0.48 | 0.20 | 0.68 | 0.15 | 0.68 | 0.55 | 0.10 | 0.63 | 0.15 | 0.31 | 0.26 | 0.30 | 0.63 | 0.55 | 0.42 |
| E | 0.43 | 0.40 | 0.09 | 1 | 0.48 | 0.38 | 0.10 | 0.58 | 0.06 | 0.58 | 0.45 | 0.00 | 0.54 | 0.06 | 0.40 | 0.17 | 0.21 | 0.54 | 0.45 | 0.33 |
| F | 0.05 | 0.08 | 0.58 | 0.48 | 1 | 0.10 | 0.38 | 0.09 | 0.42 | 0.09 | 0.03 | 0.48 | 0.05 | 0.42 | 0.89 | 0.31 | 0.27 | 0.05 | 0.03 | 0.15 |
| G | 0.04 | 0.01 | 0.48 | 0.38 | 0.10 | 1 | 0.28 | 0.19 | 0.32 | 0.19 | 0.07 | 0.37 | 0.15 | 0.32 | 0.79 | 0.21 | 0.17 | 0.15 | 0.06 | 0.05 |
| H | 0.32 | 0.29 | 0.20 | 0.10 | 0.38 | 0.28 | 1 | 0.47 | 0.04 | 0.47 | 0.35 | 0.09 | 0.43 | 0.04 | 0.51 | 0.06 | 0.10 | 0.43 | 0.34 | 0.22 |
| I | 0.15 | 0.18 | 0.68 | 0.58 | 0.09 | 0.19 | 0.47 | 1 | 0.52 | 0 | 0.12 | 0.57 | 0.04 | 0.53 | 0.99 | 0.41 | 0.37 | 0.04 | 0.12 | 0.25 |
| K | 0.36 | 0.34 | 0.15 | 0.06 | 0.42 | 0.32 | 0.04 | 0.52 | 1 | 0.52 | 0.39 | 0.05 | 0.47 | 0.00 | 0.46 | 0.10 | 0.14 | 0.47 | 0.39 | 0.27 |
| L | 0.15 | 0.18 | 0.68 | 0.58 | 0.09 | 0.19 | 0.47 | 0 | 0.52 | 1 | 0.12 | 0.57 | 0.04 | 0.52 | 0.99 | 0.41 | 0.37 | 0.04 | 0.12 | 0.25 |
| M | 0.02 | 0.05 | 0.55 | 0.45 | 0.03 | 0.07 | 0.35 | 0.12 | 0.39 | 0.12 | 1 | 0.44 | 0.08 | 0.39 | 0.86 | 0.28 | 0.24 | 0.08 | 0.00 | 0.12 |
| N | 0.42 | 0.39 | 0.10 | 0.00 | 0.48 | 0.37 | 0.09 | 0.57 | 0.05 | 0.57 | 0.44 | 1 | 0.53 | 0.05 | 0.41 | 0.16 | 0.20 | 0.53 | 0.44 | 0.32 |
| P | 0.11 | 0.13 | 0.63 | 0.54 | 0.05 | 0.15 | 0.43 | 0.04 | 0.47 | 0.04 | 0.08 | 0.53 | 1 | 0.47 | 0.94 | 0.37 | 0.33 | 0 | 0.08 | 0.20 |
| Q | 0.36 | 0.34 | 0.15 | 0.06 | 0.42 | 0.32 | 0.04 | 0.52 | 0.00 | 0.52 | 0.39 | 0.05 | 0.47 | 1 | 0.46 | 0.10 | 0.14 | 0.47 | 0.39 | 0.27 |
| R | 0.83 | 0.81 | 0.31 | 0.40 | 0.89 | 0.79 | 0.51 | 0.99 | 0.46 | 0.99 | 0.86 | 0.41 | 0.94 | 0.46 | 1 | 0.57 | 0.61 | 0.94 | 0.86 | 0.73 |
| S | 0.26 | 0.23 | 0.26 | 0.17 | 0.31 | 0.21 | 0.06 | 0.41 | 0.10 | 0.41 | 0.28 | 0.16 | 0.37 | 0.10 | 0.57 | 1 | 0.04 | 0.37 | 0.28 | 0.16 |
| T | 0.21 | 0.19 | 0.30 | 0.21 | 0.27 | 0.17 | 0.10 | 0.37 | 0.14 | 0.37 | 0.24 | 0.20 | 0.33 | 0.14 | 0.61 | 0.04 | 1 | 0.33 | 0.24 | 0.12 |
| V | 0.11 | 0.13 | 0.63 | 0.54 | 0.05 | 0.15 | 0.43 | 0.04 | 0.47 | 0.04 | 0.08 | 0.53 | 0 | 0.47 | 0.94 | 0.37 | 0.33 | 1 | 0.08 | 0.20 |
| W | 0.02 | 0.05 | 0.55 | 0.45 | 0.03 | 0.06 | 0.34 | 0.12 | 0.39 | 0.12 | 0.00 | 0.44 | 0.08 | 0.39 | 0.86 | 0.28 | 0.24 | 0.08 | 1 | 0.12 |
| Y | 0.09 | 0.07 | 0.42 | 0.33 | 0.15 | 0.05 | 0.22 | 0.25 | 0.27 | 0.25 | 0.12 | 0.32 | 0.20 | 0.27 | 0.73 | 0.16 | 0.12 | 0.20 | 0.12 | 1 |

Table **11**. BLOSUM62+Hydrophobicity

| | A | C | D | E | F | G | H | I | K | L | M | N | P | Q | R | S | T | V | W | Y |
|---|---|---|---|---|---|---|---|---|---|---|---|---|---|---|---|---|---|---|---|---|
| A | 4 | 0.02 | -1.47 | -0.56 | -1.94 | 0.04 | -1.67 | -0.84 | -0.63 | -0.84 | -0.97 | -1.57 | -0.88 | -0.63 | -0.16 | 1.26 | 0.21 | 0.11 | -2.97 | -1.90 |
| C | 0.02 | 9 | -2.5 | -3.59 | -1.91 | -2.98 | -2.70 | -0.81 | -2.65 | -0.81 | -0.94 | -2.60 | -2.86 | -2.65 | -2.19 | -0.76 | -0.80 | -0.86 | -1.94 | -1.92 |
| D | -1.47 | -2.5 | 6 | 2.09 | -2.41 | -0.51 | -0.79 | -2.31 | -0.84 | -3.31 | -2.44 | 1.10 | -0.36 | 0.15 | -1.69 | 0.26 | -0.69 | -2.36 | -3.44 | -2.57 |
| E | -0.56 | -3.59 | 2.09 | 5 | -2.51 | -1.61 | 0.10 | -2.41 | 1.06 | -2.41 | -1.54 | 0.00 | -0.45 | 2.06 | 0.40 | 0.17 | -0.78 | -1.45 | -2.54 | -1.66 |
| F | -1.94 | -1.91 | -2.41 | -2.51 | 6 | -2.89 | -0.61 | 0.09 | -2.57 | 0.09 | 0.03 | -2.51 | -3.94 | -2.57 | -2.10 | -1.68 | -1.72 | -0.94 | 1.03 | 3.15 |
| G | 0.04 | -2.98 | -0.51 | -1.61 | -2.89 | 6 | -1.72 | -3.80 | -1.67 | -3.80 | -2.92 | 0.37 | -1.84 | -1.67 | -1.20 | 0.21 | -1.82 | -2.84 | -1.93 | -2.94 |
| H | -1.67 | -2.70 | -0.79 | 0.10 | -0.61 | -1.72 | 8 | -2.52 | -0.95 | -2.52 | -1.64 | 1.09 | -1.56 | 0.04 | 0.51 | -0.93 | -1.89 | -2.56 | -1.65 | 2.22 |
| I | -0.84 | -0.81 | -2.31 | -2.41 | 0.09 | -3.80 | -2.52 | 4 | -2.47 | 2 | 1.12 | -2.42 | -2.95 | -2.47 | -2.00 | -1.58 | -0.62 | 3.04 | -2.87 | -0.74 |
| K | -0.63 | -2.65 | -0.84 | 1.06 | -2.57 | -1.67 | -1.95 | -2.47 | 5 | -1.47 | -0.60 | 0.05 | -0.52 | 1.00 | 2.46 | 0.10 | -0.85 | -1.52 | -2.60 | -1.72 |
| L | -0.84 | -0.81 | -3.31 | -2.41 | 0.09 | -3.80 | -2.52 | 2 | -1.47 | 4 | 2.12 | -2.42 | -2.95 | -1.47 | -1.00 | -1.58 | -0.62 | 1.04 | -1.87 | -0.74 |
| M | -0.97 | -0.94 | -2.44 | -1.54 | 0.03 | -2.92 | -1.64 | 1.12 | -0.60 | 2.12 | 5 | -1.55 | -1.91 | 0.39 | -0.13 | -0.71 | -0.75 | 1.08 | -0.99 | -0.87 |
| N | -1.57 | -2.60 | 1.10 | -2.99 | 0.48 | 1.37 | -2.90 | 0.57 | 0.05 | -2.42 | -1.55 | 6 | -1.46 | 0.05 | 0.41 | 1.16 | 0.20 | -2.46 | -3.55 | -1.67 |
| P | -0.88 | -2.86 | -0.36 | -0.45 | -3.94 | -1.84 | -1.56 | -2.95 | -0.52 | -2.95 | -1.91 | -1.46 | 7 | -0.52 | -1.05 | -0.62 | -0.66 | -2 | -3.91 | -2.79 |
| Q | -0.63 | -2.65 | 0.15 | 2.06 | -2.57 | -1.67 | 0.04 | -2.47 | 1.00 | -1.47 | 0.39 | 0.05 | -0.52 | 5 | 1.46 | -0.89 | -0.85 | -0.52 | -1.60 | -1.73 |
| R | -0.16 | -2.19 | -1.6 | 0.40 | -2.10 | -1.20 | 0.51 | -2.00 | 2.46 | -1.00 | -0.13 | 0.41 | -1.05 | 1.46 | 5 | -0.42 | -0.38 | -2.05 | -2.13 | -1.26 |
| S | 1.26 | -0.76 | 0.26 | 0.17 | -1.68 | 0.21 | -0.93 | -1.58 | 0.10 | -1.58 | -0.71 | 1.16 | -0.62 | 0.10 | -0.42 | 4 | 1.04 | -1.62 | -2.71 | -1.83 |
| T | 0.21 | -0.80 | -0.69 | -0.78 | -1.72 | -1.82 | -1.89 | -0.62 | -0.85 | -0.62 | -0.75 | 0.20 | -0.66 | -0.85 | -0.38 | 1.04 | 5 | 0.33 | -1.75 | -1.87 |
| V | 0.11 | -0.86 | -2.36 | -1.45 | -0.94 | -2.84 | -2.56 | 3.04 | -1.52 | 1.04 | 1.08 | -2.46 | -2 | -1.52 | -2.05 | -1.62 | 0.33 | 4 | -2.91 | -0.79 |
| W | -2.97 | -1.94 | -3.44 | -2.54 | 1.03 | -1.93 | -1.65 | -2.87 | -2.60 | -1.87 | -0.99 | -3.55 | -3.91 | -1.60 | -2.13 | -2.71 | -1.75 | -2.91 | 11 | 2.12 |
| Y | -1.90 | -1.92 | -2.57 | -1.66 | 3.150 | -2.94 | 2.22 | -0.74 | -1.72 | -0.74 | -0.87 | -1.67 | -2.79 | -0.73 | -1.26 | -1.83 | -1.87 | -0.79 | 2.12 | 7 |

Table **12**. Orthogonal+BLOSUM62

| | A | C | D | E | F | G | H | I | K | L | M | N | P | Q | R | S | T | V | W | Y |
|---|---|---|---|---|---|---|---|---|---|---|---|---|---|---|---|---|---|---|---|---|
| A | 5 | 0 | -2 | -1 | -2 | 0 | -2 | -1 | -1 | -1 | -1 | -2 | -1 | -1 | -1 | 1 | 0 | 0 | -3 | -2 |
| C | 0 | 10 | -3 | -4 | -2 | -3 | -3 | -1 | -3 | -1 | -1 | -3 | -3 | -3 | -3 | -1 | -1 | -1 | -2 | -2 |
| D | -2 | -3 | 7 | 2 | -3 | -1 | -1 | -3 | -1 | -4 | -3 | 1 | -1 | 0 | -2 | 0 | -1 | -3 | -4 | -3 |
| E | -1 | -4 | 2 | 6 | -3 | -2 | 0 | -3 | 1 | -3 | -2 | 0 | -1 | 2 | 0 | 0 | -1 | -2 | -3 | -2 |
| F | -2 | -2 | -3 | -3 | 7 | -3 | -1 | 0 | -3 | 0 | 0 | -3 | -4 | -3 | -3 | -2 | -2 | -1 | 1 | 3 |
| G | 0 | -3 | -1 | -2 | -3 | 7 | -2 | -4 | -2 | -4 | -3 | 0 | -2 | -2 | -2 | 0 | -2 | -3 | -2 | -3 |
| H | -2 | -3 | -1 | 0 | -1 | -2 | 9 | -3 | -1 | -3 | -2 | 1 | -2 | 0 | 0 | -1 | -2 | -3 | -2 | 2 |
| I | -1 | -1 | -3 | -3 | 0 | -4 | -3 | 5 | -3 | 2 | 1 | -3 | -3 | -3 | -3 | -2 | -1 | 3 | -3 | -1 |
| K | -1 | -3 | -1 | 1 | -3 | -2 | -2 | -3 | 6 | -2 | -1 | 0 | -1 | 1 | 2 | 0 | -1 | -2 | -3 | -2 |
| L | -1 | -1 | -4 | -3 | 0 | -4 | -3 | 2 | -2 | 5 | 2 | -3 | -3 | -2 | -2 | -2 | -1 | 1 | -2 | -1 |
| M | -1 | -1 | -3 | -2 | 0 | -3 | -2 | 1 | -1 | 2 | 6 | -2 | -2 | 0 | -1 | -1 | -1 | 1 | -1 | -1 |
| N | -2 | -3 | 1 | -3 | 0 | 1 | -3 | 0 | 0 | -3 | -2 | 7 | -2 | 0 | 0 | 1 | 0 | -3 | -4 | -2 |
| P | -1 | -3 | -1 | -1 | -4 | -2 | -2 | -3 | -1 | -3 | -2 | -2 | 8 | -1 | -2 | -1 | -1 | -2 | -4 | -3 |
| Q | -1 | -3 | 0 | 2 | -3 | -2 | 0 | -3 | 1 | -2 | 0 | 0 | -1 | 6 | 1 | -1 | -1 | -1 | -2 | -2 |
| R | -1 | -3 | -2 | 0 | -3 | -2 | 0 | -3 | 2 | -2 | -1 | 0 | -2 | 1 | 6 | -1 | -1 | -3 | -3 | -2 |
| S | 1 | -1 | 0 | 0 | -2 | 0 | -1 | -2 | 0 | -2 | -1 | 1 | -1 | 0 | -1 | 5 | 1 | -2 | -3 | -2 |
| T | 0 | -1 | -1 | -1 | -2 | -2 | -2 | -1 | -1 | -1 | -1 | 0 | -1 | -1 | -1 | 1 | 6 | 0 | -2 | -2 |
| V | 0 | -1 | -3 | -2 | -1 | -3 | -3 | 3 | -2 | 1 | 1 | -3 | -2 | -2 | -3 | -2 | 0 | 5 | -3 | -1 |
| W | -3 | -2 | -4 | -3 | 1 | -2 | -2 | -3 | -3 | -2 | -1 | -4 | -4 | -2 | -3 | -3 | -2 | -3 | 12 | 2 |
| Y | -2 | -2 | -3 | -2 | 3 | -3 | 2 | -1 | -2 | -1 | -1 | -2 | -3 | -1 | -2 | -2 | -2 | -1 | 2 | 8 |

Table **13**. PAM250+Hydrophobicity

| | A | C | D | E | F | G | H | I | K | L | M | N | P | Q | R | S | T | V | W | Y |
|---|---|---|---|---|---|---|---|---|---|---|---|---|---|---|---|---|---|---|---|---|
| A | 2 | -1.97 | 0.52 | 0.43 | -2.94 | 1.04 | -0.67 | -0.84 | -0.63 | -1.84 | -0.97 | 0.42 | 1.11 | 0.36 | -1.16 | 1.26 | 1.21 | 0.11 | -5.97 | -2.90 |
| C | -1.97 | -4 | -0.5 | -0.59 | -3.91 | -2.98 | 2.29 | -1.81 | 3.34 | -2.81 | 0.05 | 0.39 | 0.13 | 1.34 | 6.81 | 0.23 | -0.80 | -1.86 | 2.05 | -3.92 |
| D | 0.52 | -3.5 | 2 | 1.09 | -2.41 | 0.48 | 2.20 | -1.31 | 1.15 | -2.31 | -1.44 | 2.10 | 0.63 | 1.15 | 0.31 | 1.26 | 0.30 | -1.36 | -3.44 | -1.57 |
| E | 0.431 | -4.59 | 4.09 | 3 | -5.51 | 1.38 | 1.10 | -1.41 | 0.06 | -3.41 | -2.54 | 2.00 | -0.45 | 2.06 | -0.59 | 0.17 | 0.21 | -1.45 | -6.54 | -3.66 |
| F | -1.94 | 12.08 | -4.41 | -4.51 | -4 | -2.89 | -2.61 | -1.90 | -4.57 | -5.90 | -4.96 | -3.59 | -2.94 | -4.57 | -3.10 | 0.31 | -1.72 | -1.94 | -7.96 | 0.15 |
| G | 0.04 | -4.98 | 2.48 | 2.38 | -4.89 | -1 | 3.28 | -1.80 | 1.32 | -1.80 | -0.92 | 1.39 | 0.15 | 4.32 | 1.79 | -0.78 | -0.82 | -1.84 | -4.93 | -3.94 |
| H | 0.32 | -4.70 | 3.20 | 4.10 | -4.61 | 0.28 | 1 | -1.52 | 0.04 | -2.52 | -1.64 | 1.09 | -0.56 | 2.04 | -0.48 | 0.06 | 0.10 | -1.56 | -6.65 | -3.77 |
| I | 1.15 | -2.81 | 1.68 | 0.58 | -4.90 | 5.19 | -1.52 | -3 | -1.47 | -4 | -2.87 | 0.58 | 0.04 | -0.47 | -2.00 | 1.41 | 0.37 | -0.95 | -6.87 | -4.74 |
| K | -0.632 | -2.65 | 1.15 | 1.06 | -1.57 | -1.67 | 6.04 | -1.47 | 0 | -1.47 | -1.60 | 2.05 | 0.47 | 3.00 | 2.46 | -0.89 | -0.85 | -1.52 | -2.60 | 0.27 |
| L | -0.84 | -1.81 | -1.31 | -1.41 | 1.09 | -2.80 | -1.52 | 5 | -1.47 | 2 | 2.12 | -1.42 | -1.95 | -1.47 | -1.00 | -0.58 | 0.37 | 4.04 | -4.87 | -0.74 |
| M | -1.97 | -5.94 | -3.44 | -2.54 | 2.03 | -3.92 | -1.64 | 2.12 | -2.60 | 6.12 | 4 | -2.55 | -2.91 | -1.60 | -2.13 | -2.71 | -1.75 | 2.08 | -1.99 | -0.87 |
| N | -0.57 | -4.60 | 0.10 | 0.00 | -4.51 | -1.62 | 0.09 | -1.42 | 5.05 | -2.42 | 0.44 | 1 | -0.46 | -0.94 | 3.41 | 0.16 | 0.20 | -1.46 | -2.55 | -3.67 |
| P | -0.88 | -4.86 | -2.36 | -1.45 | 0.05 | -2.84 | -1.56 | 2.04 | 0.47 | 4.04 | 6.08 | -1.46 | -2 | -0.52 | 0.94 | -1.62 | -0.66 | 2 | -3.91 | -1.79 |
| Q | -2.63 | -3.69 | -5.84 | -4.93 | 9.42 | -4.67 | -1.95 | 1.52 | -4.99 | 2.52 | 0.35 | -2.94 | -4.52 | -5 | -3.53 | -2.89 | -2.85 | -0.52 | 0.39 | 7.27 |
| R | 1.83 | -2.19 | -0.69 | -0.59 | -4.10 | 0.79 | 0.513 | -1.00 | -0.53 | -2.00 | -1.13 | 0.41 | 6.94 | 0.46 | 0 | 1.57 | 0.61 | -0.05 | -5.13 | -4.26 |
| S | 1.26 | 0.23 | 0.26 | 0.17 | -2.68 | 1.21 | -0.93 | -0.58 | 0.10 | -2.58 | -1.71 | 1.16 | 1.37 | -0.89 | 0.57 | 2 | 1.04 | -0.62 | -1.71 | -2.83 |
| T | 1.21 | -1.80 | 0.30 | 0.21 | -2.72 | 0.17 | -0.89 | 0.37 | 0.14 | -1.62 | -0.75 | 0.20 | 0.33 | -0.85 | -0.38 | 1.04 | 3 | 0.33 | -4.75 | -2.87 |
| V | -5.88 | -7.82 | -6.36 | -6.45 | 0.05 | -6.84 | -2.56 | -4.95 | -2.52 | -1.95 | -3.91 | -3.46 | -6 | -4.52 | 2.94 | -1.62 | -4.66 | -6 | 17.08 | 0.20 |
| W | -2.97 | 0.05 | -3.44 | -3.54 | 7.03 | -4.93 | 0.34 | -0.87 | -3.60 | -0.87 | -1.99 | -1.55 | -4.91 | -3.60 | -3.13 | -2.71 | -2.75 | -1.91 | 0 | 10.12 |
| Y | 0.09 | -1.92 | -1.57 | -1.66 | -0.84 | -0.94 | -1.77 | 4.25 | -1.72 | 2.25 | 2.12 | -1.67 | -0.79 | -1.73 | -1.26 | -0.83 | 0.12 | 4.20 | -5.87 | -2 |

Table **14.** Orthogonal+PAM250

| | A | C | D | E | F | G | H | I | K | L | M | N | P | Q | R | S | T | V | W | Y |
|---|---|---|---|---|---|---|---|---|---|---|---|---|---|---|---|---|---|---|---|---|
| A | 3 | -2 | 0 | 0 | -3 | 1 | -1 | -1 | -1 | -2 | -1 | 0 | 1 | 0 | -2 | 1 | 1 | 0 | -6 | -3 |
| R | -2 | -3 | -1 | -1 | -4 | -3 | 2 | -2 | 3 | -3 | 0 | 0 | 0 | 1 | 6 | 0 | -1 | -2 | 2 | -4 |
| N | 0 | -4 | 3 | 1 | -3 | 0 | 2 | -2 | 1 | -3 | -2 | 2 | 0 | 1 | 0 | 1 | 0 | -2 | -4 | -2 |
| D | 0 | -5 | 4 | 4 | -6 | 1 | 1 | -2 | 0 | -4 | -3 | 2 | -1 | 2 | -1 | 0 | 0 | -2 | -7 | -4 |
| C | -2 | 12 | -5 | -5 | -3 | -3 | -3 | -2 | -5 | -6 | -5 | -4 | -3 | -5 | -4 | 0 | -2 | -2 | -8 | 0 |
| Q | 0 | -5 | 2 | 2 | -5 | 0 | 3 | -2 | 1 | -2 | -1 | 1 | 0 | 4 | 1 | -1 | -1 | -2 | -5 | -4 |
| E | 0 | -5 | 3 | 4 | -5 | 0 | 2 | -2 | 0 | -3 | -2 | 1 | -1 | 2 | -1 | 0 | 0 | -2 | -7 | -4 |
| G | 1 | -3 | 1 | 0 | -5 | 5 | -2 | -2 | -2 | -4 | -3 | 0 | 0 | -1 | -3 | 1 | 0 | -1 | -7 | -5 |
| H | -1 | -3 | 1 | 1 | -2 | -2 | 6 | -2 | 1 | -2 | -2 | 2 | 0 | 3 | 2 | -1 | -1 | -2 | -3 | 0 |
| I | -1 | -2 | -2 | -2 | 1 | -3 | -2 | 5 | -2 | 3 | 2 | -2 | -2 | -2 | -2 | -1 | 0 | 4 | -5 | -1 |
| L | -2 | -6 | -4 | -3 | 2 | -4 | -2 | 2 | -3 | 6 | 5 | -3 | -3 | -2 | -3 | -3 | -2 | 2 | -2 | -1 |
| K | -1 | -5 | 0 | 0 | -5 | -2 | 0 | -2 | 5 | -3 | 0 | 2 | -1 | -1 | 3 | 0 | 0 | -2 | -3 | -4 |
| M | -1 | -5 | -3 | -2 | 0 | -3 | -2 | 2 | 0 | 4 | 6 | -2 | -1 | -1 | 0 | -2 | -1 | 2 | -4 | -2 |
| F | -3 | -4 | -6 | -5 | 9 | -5 | -2 | 1 | -5 | 2 | 0 | -3 | -5 | -4 | -4 | -3 | -3 | -1 | 0 | 7 |
| P | 1 | -3 | -1 | -1 | -5 | 0 | 0 | -2 | -1 | -3 | -2 | 0 | 6 | 0 | 1 | 1 | 0 | -1 | -6 | -5 |
| S | 1 | 0 | 0 | 0 | -3 | 1 | -1 | -1 | 0 | -3 | -2 | 1 | 1 | -1 | 0 | 3 | 1 | -1 | -2 | -3 |
| T | 1 | -2 | 0 | 0 | -3 | 0 | -1 | 0 | 0 | -2 | -1 | 0 | 0 | -1 | -1 | 1 | 4 | 0 | -5 | -3 |
| W | -6 | -8 | -7 | -7 | 0 | -7 | -3 | -5 | -3 | -2 | -4 | -4 | -6 | -5 | 2 | -2 | -5 | -5 | 17 | 0 |
| Y | -3 | 0 | -4 | -4 | 7 | -5 | 0 | -1 | -4 | -1 | -2 | -2 | -5 | -4 | -4 | -3 | -3 | -2 | 1 | 10 |
| V | 0 | -2 | -2 | -2 | -1 | -1 | -2 | 4 | -2 | 2 | 2 | -2 | -1 | -2 | -2 | -1 | 0 | 4 | -6 | -1 |

### 2.3. Sliding Window Protocol

The encoding schemes discussed above give the input data in matrix format with 20 columns and number of rows corresponding to the length of the sequence. This input data is subdivided as per the sliding window protocol. This protocol helps in designing dynamic neural network architecture in which the number of units in the input layer is created corresponding to window size. Window size is an odd number so that the amino acid at the centre of the window is predicted. The method uses window size ranging from 3, 5, 7,9,11,13,15,17, and 19.

As per this, for the given input data of matrix with hundred rows, twenty columns and a chosen window size, which we assume as three, an input layer of three units is created in the network. The first three rows of the matrix are supplied as input to these units. Subsequently, as the window slides one per amino acid, collection of three rows of the matrix is supplied iteratively as shown in Figure 5. Similarly, a neural network architecture with number of units in the input layer as 5, 7,9,11,13,15,17 and 19 are created corresponding to the respective window sizes.

### 2.4. Number of units in the hidden layer

The proposed methodology uses a positive integer from 1 to 20 as distinct values for the number of units in the hidden layer. Based on the selected number, neural network architecture with that many units in the hidden layer is created

### 2.5. Types of learning algorithms

The method uses different types of learning algorithms. Different classes of learning algorithms like conjugate gradient algorithms, heuristic algorithms and quasi- Newton algorithms are used in the methodology.

#### 2.5.1 Conjugate gradient algorithms

Conjugate gradient algorithms are a class of learning algorithms in which the search process is undertaken along the conjugate directions. These search directions are periodically reset to the negative of the gradient. The standard reset point occurs when the number of iterations is equal to the number of network weights. Such algorithms require less storage space and converge faster. They are good for networks with large number of connections [53] [73].

The work uses four conjugate gradient algorithms. They are Scale gradient Conjugate Back Propagation (SCG), Conjugate gradient Back Propagation with Polak–Riebre Updates (CGP), Conjugate gradient Back Propagation with Fletcher-Reeves Updates (CGF) and Conjugate gradient Back Propagation with Powell-Beale Restarts (CGB). Scaled Conjugate Gradient uses the mechanism of step size scaling to avoid line search per iteration. Though it requires more iterations, it uses less number of computations and is a fast converging algorithm [99]. Conjugate Gradient Back Propagation with Fletcher-Reeves Updates learning algorithm uses the ratio of the norm squared of the current gradient to the norm squared of the previous gradient. Such algorithms are usually faster than other similar algorithms [45]. Conjugate Gradient Back Propagation with Polak-Riebre Updates uses the ratio of the inner product of the previous change in the gradient with the current gradient to the norm squared of the previous gradient. It requires slightly larger storage than Fletcher-Reeves [36]. Conjugate Gradient Back Propagation with Powell- Beale Restarts uses a reset method proposed by Powell [110] based on the one suggested by Beale [14]. According to this, the restart takes place if there is very small orthogonality left between the current gradient and the previous gradient. It requires more storage space than the Polak-Riebre.

#### 2.5.2 Heuristic algorithms

Heuristic algorithms are the learning algorithms, which use search and problem specific information for better performance. The proposed work uses Resilient Back Propagation and Variable Learning algorithm.

Resilient Back Propagation (RBP) training algorithm uses only the sign and not the magnitude of the derivative of the error function for the weight update [113]. The update value for the weight is increased by a factor if the derivative of the error function has the same sign for two successive iterations and is decreased otherwise. It remains the same if the derivative is zero. Variable Learning Rate (VLR) algorithm uses a separate mechanism for the learning rate as used in standard Steepest Descent Learning algorithm. It adopts an adaptive learning rate, which incorporates a momentum factor. Thus the learning algorithm becomes more responsive to the fluctuations of local error [132].

#### 2.5.3 Quasi-Newton algorithms

Newton method is a class of hill-climbing technique that looks for a stationary point of an error function. This necessitates the creation of a matrix called Hessian matrix whose entries are the second derivatives of the error function with respect to the weights at current positions. However, calculation of Hessian matrix requires more space and time. Quasi-Newton method uses an approximation, instead of the actual calculation, of these second order derivatives.

The work uses two types of quasi-Newton algorithms they are Broyden–Fletcher–Goldfarb–Shanno (BFGS) algorithm and One-Step Secant (OSS) algorithm. BFGS uses an approximation to the second order derivative as suggested by [23][46][51][122]. It requires more storage space but converges faster than the conjugate gradient methods. One-Step Secant Algorithm [13] uses identity matrix as the Hessian matrix for the previous iteration. This reduces the requirement of more storage space, which is the limitation of BFGS algorithm. However, OSS requires a little more storage space than the conjugate gradient algorithms.

Table 15 gives the learning algorithms and the corresponding formula used in the proposed method.

Table **15**. List of Algorithms

| **Algorithm** | **Weight Adaptation** |
|---|---|
| Gradient descent with momentum and adaptive learning rate (VLR) | $\Delta w_k = p.\Delta w_{k-1} + \alpha.p.\frac{\Delta E_k}{\Delta w_k}$ |
| Resilient BP | $\Delta w_k = -sign\left(\frac{\Delta E_k}{\Delta w_k}\right)\Delta_k$ |
| Scale conjugate gradient (conjugate) | $w_k = w_{k-1} + \alpha_k p_{k-1}$ |
| Fletcher-Reeves (conjugate) | $p_0 = -g_0$<br>$\Delta w_k = \alpha_k p_k$<br>$p_k = -g_k + \beta_k p_{k-1}$<br>$\beta_k = \frac{g'_k g_k}{g'_{k-1} g_{k-1}}$ |
| Polak-Ribiere (conjugate) | $p_k = -g_k + \beta_k p_{k-1}$ $\beta_k = \frac{\Delta g'_{k-1} g_k}{g'_{k-1} g_{k-1}}$ |
| Powell-Beale restarts (conjugate) | $\lvert g'_{k-1} g_k \rvert \geq 0.2 \lVert g_k \rVert^2$ |
| quasi-Newton : BFGS | $\Delta w_k = -H'_k g_k$ |
| quasi-Newton : One Step Secant | $w_{k-1} = w_k - H_k^{-1} g_k$ |

The symbols mentioned in the formulae are described below:
$\Delta w_k$: Current vector of weights changes

$\boldsymbol{p}$**:** Input vector
$\Delta w_{k-1}$: Previous vector of weights changes
α : Learning rate

$\mathbf{E_k}$: Error function E at *k*

$\boldsymbol{\Delta E_k}$**:** Sum-of-squared-differences error function E at *k*

$\boldsymbol{\Delta_k}$ : Bias increased

$\boldsymbol{\alpha_k}$: Step size
$p_0$ : Initial search gradient
$p_k$: Current search direction
$g_k$ : is the current gradient
$\beta_k$: Constant
$p_{k-1}$: Previous search direction
$g_0$: Initial Gradient
$w_k$: Current weight vector
$w_{k+1}$: Next weight vector
$H_k$: Hessian matrix (second derivatives) matrix

$\boldsymbol{H'_k}$ **:** Inverse Hessian Matrix

$\mathbf{g'_k} \boldsymbol{g_k}$ : Norm squared of the current gradient

$\Delta \mathbf{g'_{k-1}}$ : Previous change in the gradient.

$\boldsymbol{g'_{k-1} g_{k-1}}$: Norm squared of the previous gradient.

$\mathbf{w_{k-1}}$ : Previous vector of weight change

**2.6. Performance measures**

A confusion matrix [83], which is a matrix representation that cross-tabulates observed and predicted observations [44], as depicted in Table 16, is used to derive different performance measures [83].

Table **16**. Confusion matrix

| | Classified as | | |
|---|---|---|---|
| Reference | | - | + |
| | - | TN | FP |
| | + | FN | TP |

The observed and the predicted observations are represented in the following four types:

- TP -True Positive: Number of structures correctly predicted.
- TN- True Negative: Number of structures wrongly predicted.
- FN- False Negative: Number structures under-predicted. i.e. the number of structures belong to class i but the prediction says that structure does not belong to class i.
- FP- False Positive: Number of structures over-predicted i.e. the number of structures that do not belong to class i but prediction says that it belongs to class i.

Based on this confusion matrix, five performance measures are used to estimate the prediction accuracy. They are Q3, specificity (Sp), sensitivity (Se), Mathew correlation coefficient (MCC) and accuracy. The following subsections give a brief discussion on each one of these.

The ROC plot [103] [52] is used to compare the results obtained. It is plotted by taking (1-Sp) along the X-axis and Se along the Y-axis. The best value is selected as the one, which is plotted on the upper left corner of the ROC plot.

**2.6.1 Performance measure: $Q_3$**

$Q_3$ [121] is one of the most commonly used performance measures in the protein secondary structure prediction. It refers to three state overall percentages of correctly predicted residues. This measure is defined as,

$$Q3 = \sum_{(i=H,E,C)} \frac{\text{No of residues correctly predicted}}{\text{No of residues in class}} * 100$$

(EQ.6)

Where H= α –helix, E= β-strands and C= coil/turns

**2.6.2 Sensitivity**

It describes how well a classifier classifies those observations that belong to the class. It is given by the formula

$$Se = TP/(TP + FN)$$

(EQ.7)

### 2.6.3 Specificity

It describes how well a classification task classifies those observations that do not belong to a particular class. It is given by the formula

$$Sp = TN/(TN + FP)$$

(EQ.8)

### 2.6.4. Matthew's correlation coefficient

The Matthews correlation coefficient [93] for each of the three structural classes of secondary structure is given by

$$C_{i \in \{H,E,C\}} = \frac{TP * TN - FP * FN}{\sqrt{(TP + FN)(TP + FP)(TN + FP)(TN + FN)}}$$

(EQ.9)

MCC always lies between -1 and +1. A value of -1 indicates total disagreement and +1 shows total agreement. A value of 0 for MCC shows that prediction is purely random.

### 2.6.5. Accuracy

Accuracy is defined as the proportion of the total number of predictions that were correct. It is determined using the equation:

$$Accuracy = (TP + TN)/(TP + FP + FN + TN)$$

(EQ.10)

## 2.7. Statistical measures

In order to find the type and scope of the relationship between the above mentioned performance measures, the proposed work uses the following statistical measures.

### 2.7.1. Karl's Pearson correlation coefficient

The Karl Pearson's correlation coefficient [107][106] 'r' for two sets of values xi and yi is given by

$$r = covariance(x_i, y_i)/(\sigma_x * \sigma_y)$$

(EQ.11)

Where,

$$covariance(x_i, y_i) = (\sum(x_i * y_i))/N$$

(EQ.12)

$$\sigma_x = \sqrt{\frac{(\sum x_i^2)}{N} - \left(\frac{\sum xi}{N}\right)^2}$$

(EQ.13)

$$\sigma_y = \sqrt{\frac{(\sum y_i^2)}{N} - \left(\frac{\sum yi}{N}\right)^2}$$

(EQ.14)

with N as the number of observations.

### 2.7.2. Spearman's rank correlation coefficient

The Spearman's rank correlation coefficient [125] 'R' between two sets of observations having ranks xi and yi is given by

$$R = 1 - \frac{(6\sum d_i^2)}{N(N^2 - 1)}$$

(EQ.15)

where '$d_i$' is the difference of ranks between $x_i$ and $y_i$ and N is the total number of observations .

## 2.8. Software

The software used for the experiments is Matlab Version 8.2.0.701 (R2013b). The Neural Network Toolbox Version 8.1 (R2013b) is used for the implementation of neural networks. The computer that was used to perform the experiments for model selection was an Intel(R) Core(TM) 2CPU6300@1.86GHz. The Operating System used is Microsoft Windows XP Version 5.1 (Build 2600: Service Pack 3) and 1024 MB RAM.

# 3. RESULTS AND DISCUSSIONS

The work discusses a series of experiments performed on a three layer feed forward back propagation neural network using RS126 dataset. The objective of the experiment was to find an optimized parameter set by changing the values of all parameters. Thus 12960 records, corresponding to 9 data encoding schemes, 9 window sizes, 20 numbers of hidden neurons and 8 types of learning algorithms were obtained. Of these, the BFGS learning algorithm could not generate all the data within the stipulated time and therefore the records corresponding to that were not taken into consideration for further analysis. Thus, the overall focus was to find an optimal predictive model based on the remaining 11340 records.

## 3.1. Analysis of performance measures

Figure 6 shows different values obtained by each of the performance measures for these 11,340 records. This demonstrates that Se, Sp, MCC and accuracy have similar trends. Table 17 shows the best results obtained by the model based on each of these performance measures. This reveals that the parameter set for the best result based on performance measure Se, Sp, MCC and accuracy is the same. Though this table shows only the best results, other results also show the similar trend. Based on Figure.6 and Table 17 we have chosen accuracy as a representative performance measure reflecting the behaviour of Sp, Se, and MCC. Thus the five different performance measures are effectively being reduced to two distinct performance measures $Q_3$ and accuracy.

Table **17**. Best results obtained by the model based on each of the performance measures.

| Performance Measures | Best Value | Model based on the parameter | | | |
|---|---|---|---|---|---|
| | | LA | ES | WS | HN |
| $Q_3$ | 62.43687 | SCG | BLOSUM62 | 11 | 18 |
| Accuracy | 0.781395 | OSS | BLOSUM62 | 19 | 19 |
| $S_e$ | 0.672092 | OSS | BLOSUM62 | 19 | 19 |
| $S_p$ | 0.836046 | OSS | BLOSUM62 | 19 | 19 |
| MCC | 0.508138 | OSS | BLOSUM62 | 19 | 19 |

The experimental results on 11, 340 records show a slightly divergent outcome based on the performance measure $Q_3$ and accuracy. This shows that the selection of performance measure as $Q_3$ or accuracy has a bearing on the best parameter set. In order to study the relation between these performance measures, we use Karl – Pearson correlation coefficient given in (EQ.11) and Spearman's rank correlation coefficient shown in (EQ. 15) to find the extent of relation between $Q_3$ and accuracy.

### 3.2. Karl Pearson's correlation coefficient between $Q_3$ and accuracy

Table 18 gives different values of correlation coefficient between accuracy and $Q_3$ with respect to different learning algorithms. The values oscillate between a small range of 0.961663 and 0.983973. The average value of the correlation coefficient between $Q_3$ and accuracy for the 11340 records classified based on different learning algorithms are observed as 0.973935.

Table **18**. Values of correlation coefficient between accuracy and $Q_3$ with respect to different learning algorithms

| Learning algorithms | correlation coefficient between accuracy and $Q_3$ 11340 records |
|---|---|
| CGF | 0.983973 |
| CGP | 0.980611 |
| CGB | 0.978822 |
| VLR | 0.977431 |
| OSS | 0.971583 |
| RBP | 0.963465 |
| SCG | 0.961663 |
| Correlation Coefficient | 0.973935 |

Table 19 shows distinct values of correlation coefficient between accuracy and $Q_3$ of all records classified based on encoding scheme. The observed values show that correlation coefficient ranges between 0.969326 and 0.979389. The average value of the correlation coefficient is given to be 0.976229.

Table **19.** Values of correlation coefficient between accuracy and $Q_3$ with respect to different encoding schemes

| Encoding Scheme | correlation coefficient between accuracy and $Q_3$11340 records |
|---|---|
| Orthogonal | 0.979389 |
| Orthogonal+Hydrophobicity | 0.979374 |
| BLOSUM62 | 0.977243 |
| BLOSUM62+ Orthogonal | 0.976558 |
| PAM250+ Orthogonal | 0.976283 |
| PAM250+ Hydrophobicity | 0.976124 |
| PAM250 | 0.975704 |
| Hydrophobicity | 0.970169 |
| BLOSUM62+ Hydrophobicity | 0.969326 |
| Correlation Coefficient | 0.976229 |

The correlation coefficient between accuracy and $Q_3$ of all records with respect to window size is shown in Table 20. The value of correlation coefficient for different window size is observed to be 1.

Table **20.** Values of correlation coefficient between accuracy and $Q_3$ with respect to different window sizes

| Window size | Correlation coefficient between accuracy and $Q_3$ 11340 records |
|---|---|
| 3 | 1 |
| 5 | 1 |
| 7 | 1 |
| 9 | 1 |
| 11 | 1 |
| 13 | 1 |
| 15 | 1 |
| 17 | 1 |
| 19 | 1 |
| Correlation Coefficient | 1 |

Table 21 depicts the relationship between $Q_3$ and accuracy for all data classified based on number of hidden neurons. It reveals that the value of correlation coefficient varies from 0.960741 to 0.990592 with an average value of 0.978894.

Table **21**. Values of correlation coefficient between accuracy and Q3 with respect to different Hidden neurons

| Hidden neurons | correlation coefficient between accuracy and $Q_3$11340 records |
|---|---|
| 10 | 0.990592 |
| 13 | 0.987586 |
| 20 | 0.98544 |
| 16 | 0.983743 |
| 7 | 0.982355 |
| 18 | 0.981784 |
| 19 | 0.980282 |
| 14 | 0.980116 |
| 15 | 0.97787 |
| 1 | 0.977721 |
| 2 | 0.973136 |
| 12 | 0.970144 |
| 4 | 0.966114 |
| 5 | 0.965647 |
| 8 | 0.965476 |
| 3 | 0.96427 |
| 9 | 0.964004 |
| 17 | 0.963024 |
| 6 | 0.961642 |
| 11 | 0.960741 |
| Correlation Coefficient | 0.978894 |

Table 22 looks at the effect of size of sample data taken randomly. The experiments were conducted to find the correlation coefficient between $Q_3$ and accuracy for all the 11340 records. Subsequently, the sample size was reduced by half and that many records were taken randomly. The process was continued upto sample size 88. The result shows that the values of the correlation coefficient vary between a small range and value for the full set of observations is found to be 0.978909.

Table **22**. Values of correlation coefficient between accuracy and $Q_3$ with respect to size of sample data taken randomly

| Sample data | correlation coefficient between accuracy and $Q_3$ on sample data taken randomly |
|---|---|
| 708 | 0.985134 |
| 354 | 0.983513 |
| 11340 | 0.978909 |
| 2835 | 0.977928 |
| 177 | 0.974477 |
| 5670 | 0.974285 |
| 1417 | 0.968037 |
| 88 | 0.943289 |
| Correlation Coefficient | 0.972381 |

This analysis of correlation coefficient between $Q_3$ and accuracy of data classified based on different yardsticks shows that there is a strong positive correlation between the performance measures $Q_3$ and accuracy. In other words, as the value of $Q_3$ increases, the value of accuracy also increases and vice versa.

### 3.3. Spearman’s rank correlation coefficient between $Q_3$ and accuracy

For calculating Spearman’s rank correlation coefficient between $Q_3$ and accuracy, each of the observations was given a rank ranging from 1 to 11,340 based on the respective measure. The Spearman’s rank correlation coefficient, calculated using the formula given in (EQ. 15) is found to be 0.952790932. This also shows that there is a strong and positive correlation between $Q_3$ and accuracy.

### 3.4. Relation between $Q_3$ and accuracy

The previous findings based on Karl Pearson’s correlation coefficient and Spearmen’s rank correlation coefficient shows that there is a strong, definite and positive correlation between the performance measures $Q_3$ and accuracy. Thus, these measures are interchangeable as far as the experimental results are concerned. Since accuracy was taken as the representative measure of Se, Sp and MCC, one can conclude that all these measures behave in a similar manner.

### 3.5. Selection of accuracy as the preferred performance measure

A glance through literature shows that $Q_3$ is the most preferred performance measure for secondary structure prediction. However, $Q_3$ focuses only on TP (True Positive) values whereas accuracy uses TP (True Positive), TN (True Negative), FP (False Positive) and FN (False Negative) values. This means accuracy carries more information than $Q_3$ as a performance measure. In addition, it has been established that there is a strong, positive and definite relation between accuracy and $Q_3$. So, accuracy is better positioned to gauge the trend of the effectiveness of the prediction under study with respect to $Q_3$. So, the method proposes accuracy as the preferred performance measure.

### 3.6. Optimal parameter set

By taking accuracy as the preferred performance measure, the top 10 best performing records are identified. An ROC curve is plotted in Figure 7 with these best performing records. The leftmost upper corner of the ROC curve is identified as the best performing one, which is reached by BLOSUM62 as the encoding scheme, 19 as the window size, 19 as the number of neurons in the hidden layer and OSS as the learning algorithm. These values of the parameters are proposed as the optimized parameter set for the three layer feed forward back propagation neural network.

Table 23 gives the values of Mean Squared Error (MSE) for the best performing 10 records. It shows that the MSE for the best parameter set occupies the second slot with respect to MSE. The difference in MSE of this best parameter set with the set having the best MSE value is 0.000094 only. This validates the effectiveness of the optimized parameter set.

Table **23**. Mean Squared Error (MSE) for the best performing 10 records from 11340 records

| Learning algorithm | Encoding Scheme | Window Size | Hidden Neurons | MSE |
|---|---|---|---|---|
| RBP | PAM250+Orthogonal | 19 | 18 | 0.144958 |
| OSS | BLOSUM62 | 19 | 19 | 0.145052 |
| SCG | BLOSUM62+Orthogonal | 19 | 15 | 0.145197 |
| CGB | BLOSUM62 | 19 | 19 | 0.1457 |
| OSS | BLOSUM62+Orthogonal | 17 | 17 | 0.1457 |
| CGB | PAM250+Orthogonal | 19 | 19 | 0.145723 |
| SCG | BLOSUM62 | 19 | 15 | 0.146144 |
| SCG | Orthogonal+Hydrophobicity | 13 | 17 | 0.146358 |
| CGB | PAM250+Hydrophobicity | 19 | 20 | 0.146391 |
| CGB | PAM250 | 19 | 20 | 0.146391 |

### 3.7. Generation of stabilized cluster of records using accuracy

In order to study the nature of accuracy of all the records, we arranged 11340 records in increasing order with respect to accuracy. Figure 8 shows that rate of increase of accuracy doesn't seem to be uniform. It also shows that the rate of accuracy increases substantially up to around 0.70 and the rate seems to be stabilized by then.

To identify the records where the rate of convergence of accuracy seems to be uniform, we looked at the records having accuracy more than 0.70. Figure 9A, Figure 9B, Figure 9C, Figure 9D, Figure 9E, Figure 9F, Figure 9G and Figure 9H show the trend of the records having accuracy more than 0.70, 0.71, 0.72, 0.73, 0.74, 0.75, 0.76 and 0.77 respectively. There is only one record having accuracy greater than 0.78 and so the figure showing accuracy greater than 0.78 has not been included.

A glance through these figures shows that all of them are monotonic and strictly increasing. However, the last three figures viz. Figure 9E, Figure 9G and Figure 9H show more interesting features than the rest. They are not only monotonic increasing, but they are concave and they have almost uniform rate of convergence. We focus on such records, which are monotonic and concave with uniform rate of convergence. The records contributing to these properties are the ones having accuracy not less than 0.75. These records, which are 2530 in number, are labeled as stabilized cluster of records. These records are analyzed in detail to find the effect of each of the parameters under study.

#### 3.7.1. Effect of learning algorithms on stabilized cluster of records

Table 24 gives the percentage of occurrences of each learning algorithm in the entire records and in the stabilized cluster of records. Of the total seven learning algorithms used in the study, it was observed that variable learning rate (VLR) algorithm has not appeared in the stabilized cluster of records and the contribution of CGF is minimum. The learning algorithms SCG, RBP and OSS, which collectively contribute around 60%, are the best performing learning algorithms though each of the seven algorithms has the equal contribution in the entire dataset.

Table **24**. Number and the percentage of occurrences of each learning algorithm on best performing records

| Name of the learning algorithms | Percentage(%) in 11340 records | Percentage (%) in the best performing records 2530 |
|---|---|---|
| SCG | 14.28% | 20.59289 |
| RBP | 14.28% | 19.56522 |
| OSS | 14.28% | 19.16996 |
| CGP | 14.28% | 17.62846 |
| CGB | 14.28% | 17.47036 |
| CGF | 14.28% | 5.573123 |
| VLR | 14.28% | - |

### 3.7.2. Effect of encoding scheme on stabilized cluster of records

Table 25 gives the percentage of the occurrence of each encoding scheme in the entire set of records and in stabilized cluster of records. Out of the total nine encoding schemes used in the work, it was observed that the Hydrophobicity and Blosum62 + Hydrophobicity encoding schemes are absent in the stabilized cluster of records and the contribution of orthogonal encoding scheme is minimal. It is seen that PAM, PAM+ Orthogonal and PAM + Hydrophobicity encoding schemes, which appear in around 50% of the records, are the best performing encoding schemes.

Table **25**. Number and percentage of occurrences of each encoding scheme on best performing records

| Name of the encoding schemes | Percentage(%) in 11340 records | Percentage (%) in the best performing records 2530 |
|---|---|---|
| PAM250+Orthogonal | 11.11% | 17.74704 |
| PAM250 | 11.11% | 15.81028 |
| PAM250+Hydrophobicity | 11.11% | 15.81028 |
| BLOSUM62 | 11.11% | 15.6917 |
| Orthogonal + Hydrophobicity | 11.11% | 14.94071 |
| BLOSUM62 + Hydrophobicity | 11.11% | 14.11067 |
| Orthogonal | 11.11% | 5.889328 |
| Hydrophobicity | 11.11% | 0 |
| Hydrophobicity + BLOSUM62 | 11.11% | 0 |

### 3.7.3. Effect of window size on stabilized cluster of records

Table 26 illustrates the percentage of different window sizes that appeared in the entire data set and the stabilized cluster of records. It is observed that window size 15, 17, and 19 are the best performing parameter values which appear in more than 55% of the records in the stabilized cluster. It is also noted that window size 3 and 5 did not appear and window size 7 appeared only rarely in such records.

Table **26**. Number and the percentage of occurrences of window size on best performing records

| Number of the Window Size | Percentage(%) in 11340 records | Percentage (%) in the best performing records 2530 |
|---|---|---|
| 17 | 11.11% | 19.52569 |
| 19 | 11.11% | 18.7747 |
| 15 | 11.11% | 18.22134 |
| 13 | 11.11% | 16.56126 |
| 11 | 11.11% | 14.50593 |
| 9 | 11.11% | 11.81818 |
| 7 | 11.11% | 4.189723 |
| 3 | 11.11% | 0 |
| 5 | 11.11% | 0 |

### 3.7.4. Effect of hidden neurons on stabilized cluster of records

Of the total 20 neurons used in the hidden layer, 19 is the one which appeared maximum number of times in the best performing records as shown in Table 27. It also shows that lower number of neurons has lesser number of occurrences in the stabilized cluster of records though they have equal contribution of 5% each in the entire dataset.

Table **27**. Number and the percentage of occurrences of hidden neurons on best performing records

| Number of hidden neurons | Percentage(%) in 11340 records | Percentage (%) in the best performing records 2530 |
|---|---|---|
| 19 | 5% | 8.498024 |
| 13 | 5% | 7.905138 |
| 17 | 5% | 7.667984 |
| 11 | 5% | 7.43083 |
| 20 | 5% | 7.391304 |
| 12 | 5% | 7.233202 |
| 16 | 5% | 7.114625 |
| 15 | 5% | 7.035573 |
| 18 | 5% | 6.561265 |
| 9 | 5% | 6.521739 |
| 14 | 5% | 6.442688 |
| 10 | 5% | 6.007905 |
| 8 | 5% | 5.375494 |
| 7 | 5% | 4.703557 |
| 6 | 5% | 4.426877 |
| 5 | 5% | 3.399209 |
| 4 | 5% | 3.201581 |
| 3 | 5% | 1.185771 |
| 2 | 5% | 0.158103 |
| 1 | 5% | 0 |

### 3.8. Comparative study of proposed work

Table 28 gives a comparative study of proposed work with respect to similar works on secondary structure prediction using multi layer feed forward networks.

Table **28**. Comparative study

| Work by | ANN architecture | No of Parameters changed | Data Set used and type of data | Performance measures used | Relation between measures verified | Best Parameter and its Performance measure | Size of the search space | Analysis of Neighbourhood of best parameter |
|---|---|---|---|---|---|---|---|---|
| [111] | Multilayer Back propagation (two cascade networks) | Hidden layers, window Size, Encoding scheme | From Kabsch and Sander: Similar types of haemoglobin (106). | $Q_3$ and MCC, 64.3% | No | Window size 13, Hidden units 40, Second order conjunctive coding scheme | 200 | No |
| [65] | Feed forward Back Propagation | Window Size Hidden Neurons, Encoding Scheme | 62 proteins from Kabsch and Sander | $Q_3$ , 63% | No | Window Size 17, 0-Hidden-unit network, Binary encoding | 100 | No |
| [98] | Multi-layer networks trained by the Gradient Back Propagation algorithm | Window size, Hidden Layer | Kabsch and Sanders training set of 41 proteins with no homology and a test set of 19 proteins with homologies | $Q_3$, Correlation Coefficient , 58.77% | No | Window size =17 | 110 | No |
| [47] | Feed-Forward | All | Kabsch and Sander | Learning rate | No | Window size 7, | 01 | No |
| | Network | parameters are fixed. | | $Q_P(Q_L)$ , 60.67% | | Hidden layer processing unit 10, Binary encoding | | |
| [115] | Two-layered feed-forward neural network (combination of three levels of networks) | Encoding scheme | non-redundant data base of 130 protein | $Q_3$, MCC, Reliability index70.8% | No | Window 17, Multiple sequence alignment (sequence profiles) | 02 | No |
| [116] | Feed forward neural network (Two-level network cascade ) | All parameters fixed | 130 chains of water soluble globular proteins. Less than 25% pairwise similarity for length >80 used for training and testing method | $Q_3$, 69.7% | No | Window size 17, Sequence profile in binary encoding | 01 | No |
| [25] | Feed-Forward Network | Window size Hidden layer sizes | 318 chains of high resolution from PDB Non-homologus protein chain | $Q_3$, 67.0% | No | Window size 15, Hidden layer 8 , Sequence Profile encoding scheme | 44 | No |
| [136] | Feed forward Back Propagation (two networks) | Window size ,Hidden units | 382 proteins from CB396 set for training and 115 protein chains from RS126 set for testing were used. Non-homologus protein chains | $Q_{3,}$ 67.45% | No | Window size =15 Hidden units =75, Binary encoding scheme | 60 | No |
| [9] | Feed Forward Supervised learning and back propagation error algorithm | Number of Neurons in the hidden layer, Number of training epoch | CB513 Non-homologous | $Q_3$ , 62.72% | No | Window size 19. Hidden neurons 5, epoch 4000 | 144 | No |
| [100] | Feed forward architecture built in java named Java Object Oriented Neural Engine (JOONE) | All parameters fixed | 20 Proteins from PDB | $Q_{3,}$ Helix prediction 71% and Sheet Prediction 65% | No | Training pattern 4980, epochs 10000, learning rate 0.9,Momentum 0.1 | 01 | No |
| [54] | Two-level back propagation neural network | Encoding scheme | 36 Non-homologous protein from Kabsch and Sander | $Q_{3\,,}$ 71.03% | No | Window size 13, hidden layer neuron 26, Profile and Orthogonal encoding | 05 | No |
| [35] | Multilayer Preceptron (Radial Basis Function) Two Classifier | All parameters are fixed. | Six proteins from PDB | performance goal of around $10^{-4}$ | No | Hidden layers 4 Alphanumeric encoding | 01 | No |
| [2] | Two- Stage Feed Forward Network | All parameters fixed | RS126 Non-homologus | $Q_3$, MCC 75.22% | No | Window 13, Neurons in the hidden layer 25, PSSM encoding scheme | 01 | No |
| [1] | Feed Forward neural network with Back propagation | All parameter fixed | ccPDB dataset | $Q_3$, total error in prediction =0.0092 | No | Window size 3, Fixed binary bit encoding | 01 | No |
| [75] | Feed Forward | Binary Classifiers | 60 proteins form CB513 | $Q_3$ , 75.63% | No | - | - | No |
| [38] | Three layer feed forward network | All parameter fixed | PDB, RS126, Non-homologous CB513 Non-homologous | $Q_3$ , time | No | - | - | No |
| Propo-sed Work | Three layer Feed forward Back Propagation | Encoding scheme, window size, hidden neurons, Learning algorithm | RS126 Non-homologous | $Q_3$, Se, Sp, accuracy, MCC | Yes | Encoding scheme BLOSUM, Learning Algorithm = OSS, Window Size 19, Hidden neurons =19 | 12960 | Yes |

It discusses 17 related works on seven properties. The properties discussed are

- No of Parameters changed
- Data type and dataset used
- Performance measures used
- Relation between performance measures
- Best Parameter and its Performance measure
- Size of the search space
- Analysis of Neighbourhood of best parameter

### 3.8.1. No of Parameters changed

Of the sixteen works used in the comparison, seven works have the fixed parameter sets, three works have only one parameter, four works had two parameters and two works have three parameters. The proposed method uses four parameters

### 3.8.2. Data type and dataset used

Most of the works use non-homologous data. The various dataset used are Kabsch Sander, PDB, CB396 and CB513. The proposed work uses RS126 non-homologous dataset.

### 3.8.3. Performance measures used

A look at the table shows that ten works use only one performance measure, three works use two performance measures, one work uses three performance measures and no work uses more than three performance measures. Of the performance measures used in the seventeen works, $Q_3$ appears in nine works, either independently or along with other performance measures, predominantly with MCC. The proposed work uses five performance measures, which include $Q_3$. The other measures used are MCC, Sp, Se and accuracy.

### 3.8.4. Relation between performance measures

Though there are four works, which use more than one performance measure, none of them try to establish any relationship between the performance measures. However, the proposed work, which uses five performance measures, graphically shows that MCC, Sp, Se and accuracy are interrelated and behaved in a uniform manner. By taking accuracy as the representation of these, Karl's Pearson correlation coefficient and Spearman's rank correlation coefficient are used to show that there is a strong and positive correlation between $Q_3$ and other four measures.

### 3.8.5. Best Parameter and its Performance measure

Each of the works has given its best parameter though the number of such parameters is different. The proposed work offers OSS as the learning algorithm, BLOSUM62 as the encoding scheme, 19 as the window size and 19 as the number of neurons in the hidden layer as the optimized parameter set. The values of the performance measures range from 62- 75%, $Q_3$ being the performance measure in most of the cases. The proposed work provides accuracy as the performance measure with 78.1% as its value.

### 3.8.6. Size of the search space

The works compared use a small search space with highest size of the search space being 200. The proposed method works on a search space whose size is 12960, which is substantially large with respect to the complexity of search spaces of similar works.

### 3.8.7. Analysis of neighborhood of best parameter

The works compared in the study do not perform any analysis of records whose behavior is almost that of the best parameter. They offer values for the parameters whose performance measure is the highest. The proposed work goes beyond finding the best parameter to obtain a stabilized cluster of records. This cluster contains 2530 records in the neighborhood of best parameter set. All these records in this stabilized cluster of records show that they are monotonic increasing and concave with uniform rate of convergence.

## 4. CONCLUSION

The paper proposes a methodology for secondary structure prediction of proteins using three layer feed forward back propagation neural network. It uses nine encoding schemes, nine different window sizes changing from three to nineteen, twenty neurons for the hidden layer and eight different learning algorithms. After performing exhaustive experiments, the performance was measured by different performance measures like $Q_3$, specificity, sensitivity, Mathew correlation coefficient and accuracy. Since there were some variations of the values of performance measures, a detailed analysis on the best performing records were done. It shows that the parameter set consisting of nineteen as window size, nineteen as the number of neurons in the hidden layer, BLOSUM62 as the encoding scheme and one step secant as the learning algorithm gives the optimal results, independent of the type of the performance measure used. It also gives a stabilized cluster of records where they are monotonic increasing and concave with uniform rate of convergence.

## CONFLICT OF INTEREST

The authors confirm that this article content has no conflict of interest.

## ACKNOWLEDGEMENTS

The authors acknowledge the technical support offered by Mr. Sanjay Pardhi and Mr. SubodhDeolekar